\documentstyle[12pt]{article}
\advance \topmargin by -\headheight
\advance \topmargin by -\headsep

\evensidemargin \oddsidemargin
\begin{document}

\topmargin -.6in

\newcommand{\sect}[1]{\setcounter{equation}{0}\section{#1}}
\renewcommand{\theequation}{\thesection.\arabic{equation}}
\relax

\def\rf#1{(\ref{eq:#1})}
\def\lab#1{\label{eq:#1}}
\def\nonu{\nonumber}
\def\br{\begin{eqnarray}}
\def\er{\end{eqnarray}}
\def\be{\begin{equation}}
\def\ee{\end{equation}}
\def\eq{\!\!\!\! &=& \!\!\!\! }
\def\foot#1{\footnotemark\footnotetext{#1}}
\def\lb{\lbrack}
\def\rb{\rbrack}
\def\llangle{\left\langle}
\def\rrangle{\right\rangle}
\def\blangle{\Bigl\langle}
\def\brangle{\Bigr\rangle}
\def\llb{\left\lbrack}
\def\rrb{\right\rbrack}
\def\Blb{\Bigl\lbrack}
\def\Brb{\Bigr\rbrack}
\def\lcurl{\left\{}
\def\rcurl{\right\}}
\def\({\left(}
\def\){\right)}
\def\v{\vert}                     
\def\bv{\bigm\vert}               
\def\Bgv{\;\Bigg\vert}            
\def\bgv{\bigg\vert}              
\def\lskip{\vskip\baselineskip\vskip-\parskip\noindent}
\def\mskp{\par\vskip 0.3cm \par\noindent}
\def\sskp{\par\vskip 0.15cm \par\noindent}
\def\bc{\begin{center}}
\def\ec{\end{center}}
\def\Lbf#1{{\Large {\bf {#1}}}}
\def\lbf#1{{\large {\bf {#1}}}}
\relax

\def\tr{\mathop{\rm tr}}                  
\def\Tr{\mathop{\rm Tr}}                  
\newcommand\partder[2]{{{\partial {#1}}\over{\partial {#2}}}}
\newcommand\funcder[2]{{{\delta {#1}}\over{\delta {#2}}}}
\newcommand\Bil[2]{\Bigl\langle {#1} \Bigg\vert {#2} \Bigr\rangle}  
\newcommand\bil[2]{\left\langle {#1} \bigg\vert {#2} \right\rangle} 
\newcommand\me[2]{\langle {#1}\vert {#2} \rangle} 

\newcommand\sbr[2]{\left\lbrack\,{#1}\, ,\,{#2}\,\right\rbrack} 
\newcommand\Sbr[2]{\Bigl\lbrack\,{#1}\, ,\,{#2}\,\Bigr\rbrack} 
\newcommand\pbr[2]{\{\,{#1}\, ,\,{#2}\,\}}       
\newcommand\Pbr[2]{\Bigl\{ \,{#1}\, ,\,{#2}\,\Bigr\}}  
\newcommand\pbbr[2]{\lcurl\,{#1}\, ,\,{#2}\,\rcurl}  

\def\a{\alpha}
\def\b{\beta}
\def\c{\chi}
\def\d{\delta}
\def\D{\Delta}
\def\eps{\epsilon}
\def\vareps{\varepsilon}
\def\g{\gamma}
\def\G{\Gamma}
\def\grad{\nabla}
\def\h{{1\over 2}}
\def\l{\lambda}
\def\L{\Lambda}
\def\m{\mu}
\def\n{\nu}
\def\ov{\over}
\def\om{\omega}
\def\O{\Omega}
\def\p{\phi}
\def\P{\Phi}
\def\pa{\partial}
\def\tpa{{\tilde \partial}}
\def\pr{\prime}
\def\ra{\rightarrow}
\def\lra{\longrightarrow}
\def\s{\sigma}
\def\S{\Sigma}
\def\t{\tau}
\def\th{\theta}
\def\Th{\Theta}
\def\z{\zeta}
\def\ti{\tilde}
\def\wti{\widetilde}
\def\one{\hbox{{1}\kern-.25em\hbox{l}}}

\def\cA{{\cal A}}
\def\cB{{\cal B}}
\def\cC{{\cal C}}
\def\cD{{\cal D}}
\def\cE{{\cal E}}
\def\cF{{\cal F}}
\def\cG{{\cal G}}
\def\cH{{\cal H}}
\def\cL{{\cal L}}
\def\cM{{\cal M}}
\def\cN{{\cal N}}
\def\cO{{\cal O}}
\def\cP{{\cal P}}
\def\cQ{{\cal Q}}
\def\cR{{\cal R}}
\def\cS{{\cal S}}
\def\cU{{\cal U}}
\def\cV{{\cal V}}
\def\cW{{\cal W}}
\def\cY{{\cal Y}}

\def\phanta{\phantom{aaaaaaaaaaaaaaa}}
\def\phantb{\phantom{aaaaaaaaaaaaaaaaaaaaaaaaa}}
\def\phantc{\phantom{aaaaaaaaaaaaaaaaaaaaaaaaaaaaaaaaaaa}}

 \def\Winf{{\bf W_\infty}}               
\def\Win1{{\bf W_{1+\infty}}}           
\def\hWinf{{\bf {\hat W}_\infty}}       
\def\PsDA{\Psi{\cal DO}}
\def\Intres{\int dx\, {\rm Res} \; }

\def\KP3{{\bf KP_{2+1}}}
\def\mKP3{{\bf mKP_{2+1}}}
\def\KPm{{\bf (m)KP_{2+1}}}
\def\KPt{{\bf KP_{1+1}}}
\def\mKPt{{\bf mKP_{1+1}}}

\newcommand{\nit}{\noindent}
\newcommand{\ct}[1]{\cite{#1}}
\newcommand{\bi}[1]{\bibitem{#1}}
\newcommand\PRL[3]{{\sl Phys. Rev. Lett.} {\bf#1} (#2) #3}
\newcommand\NPB[3]{{\sl Nucl. Phys.} {\bf B#1} (#2) #3}
\newcommand\NPBFS[4]{{\sl Nucl. Phys.} {\bf B#2} [FS#1] (#3) #4}
\newcommand\CMP[3]{{\sl Commun. Math. Phys.} {\bf #1} (#2) #3}
\newcommand\PRD[3]{{\sl Phys. Rev.} {\bf D#1} (#2) #3}
\newcommand\PLA[3]{{\sl Phys. Lett.} {\bf #1A} (#2) #3}
\newcommand\PLB[3]{{\sl Phys. Lett.} {\bf #1B} (#2) #3}
\newcommand\JMP[3]{{\sl J. Math. Phys.} {\bf #1} (#2) #3}
\newcommand\PTP[3]{{\sl Prog. Theor. Phys.} {\bf #1} (#2) #3}
\newcommand\SPTP[3]{{\sl Suppl. Prog. Theor. Phys.} {\bf #1} (#2) #3}
\newcommand\AoP[3]{{\sl Ann. of Phys.} {\bf #1} (#2) #3}
\newcommand\RMP[3]{{\sl Rev. Mod. Phys.} {\bf #1} (#2) #3}
\newcommand\PR[3]{{\sl Phys. Reports} {\bf #1} (#2) #3}
\newcommand\FAP[3]{{\sl Funkt. Anal. Prilozheniya} {\bf #1} (#2) #3}
\newcommand\FAaIA[3]{{\sl Functional Analysis and Its Application} {\bf #1}
(#2) #3}
\def\InvM#1#2#3{{\sl Invent. Math.} {\bf #1} (#2) #3}
\newcommand\LMP[3]{{\sl Letters in Math. Phys.} {\bf #1} (#2) #3}
\newcommand\IJMPA[3]{{\sl Int. J. Mod. Phys.} {\bf A#1} (#2) #3}
\newcommand\TMP[3]{{\sl Theor. Mat. Phys.} {\bf #1} (#2) #3}
\newcommand\JPA[3]{{\sl J. Physics} {\bf A#1} (#2) #3}
\newcommand\JSM[3]{{\sl J. Soviet Math.} {\bf #1} (#2) #3}
\newcommand\MPLA[3]{{\sl Mod. Phys. Lett.} {\bf A#1} (#2) #3}
\newcommand\JETP[3]{{\sl Sov. Phys. JETP} {\bf #1} (#2) #3}
\newcommand\JETPL[3]{{\sl  Sov. Phys. JETP Lett.} {\bf #1} (#2) #3}
\newcommand\PHSA[3]{{\sl Physica} {\bf A#1} (#2) #3}
\newcommand\PHSD[3]{{\sl Physica} {\bf D#1} (#2) #3}

\newcommand\hepth[1]{{\sl hep-th/#1}}

\def\bb{{\bar  B}}
\def\bom{{\bar \om}}
\newcommand\ttmat[9]{\left(\begin{array}{ccc}  
{#1} & {#2} & {#3} \\ {#4} & {#5} & {#6} \\
{#7} & {#8} & {#9} \end{array} \right)}
\newcommand\thrcol[3]{\left(\begin{array}{c}  
{#1} \\ {#2} \\ {#3} \end{array} \right)}
\def\AM#1{A^{(M)}_{#1}}
\def\BM#1{B^{(M)}_{#1}}
\def\faa{Fa\'a di Bruno~}
\def\sj{{\jmath}}
\def\bsj{{\bar \jmath}}

\newcommand\sumi[1]{\sum_{#1}^{\infty}}   
\newcommand\fourmat[4]{\left(\begin{array}{cc}  
{#1} & {#2} \\ {#3} & {#4} \end{array} \right)}
\newcommand\twocol[2]{\left(\begin{array}{cc}  
{#1} \\ {#2} \end{array} \right)}

\def\mark{\noindent{\bf Remark.}\quad}
\def\prop{\noindent{\bf Proposition.}\quad}
\def\theor{\noindent{\bf Theorem.}\quad}
\def\name{\noindent{\bf Definition.}\quad}
\def\exam{\noindent{\bf Example.}\quad}
\def\proof{\noindent{\bf Proof.}\quad}

\font\numbers=cmss12
\font\upright=cmu10 scaled\magstep1
\def\stroke{\vrule height8pt width0.4pt depth-0.1pt}
\def\topfleck{\vrule height8pt width0.5pt depth-5.9pt}
\def\botfleck{\vrule height2pt width0.5pt depth0.1pt}
\def\Zmath{\vcenter{\hbox{\numbers\rlap{\rlap{Z}\kern 0.8pt\topfleck}\kern
2.2pt
                   \rlap Z\kern 6pt\botfleck\kern 1pt}}}
\def\Qmath{\vcenter{\hbox{\upright\rlap{\rlap{Q}\kern
                   3.8pt\stroke}\phantom{Q}}}}
\def\Nmath{\vcenter{\hbox{\upright\rlap{I}\kern 1.7pt N}}}
\def\Cmath{\vcenter{\hbox{\upright\rlap{\rlap{C}\kern
                   3.8pt\stroke}\phantom{C}}}}
\def\Rmath{\vcenter{\hbox{\upright\rlap{I}\kern 1.7pt R}}}
\def\IZ{\ifmmode\Zmath\else$\Zmath$\fi}
\def\IQ{\ifmmode\Qmath\else$\Qmath$\fi}
\def\IN{\ifmmode\Nmath\else$\Nmath$\fi}
\def\IC{\ifmmode\Cmath\else$\Cmath$\fi}
\def\IR{\ifmmode\Rmath\else$\Rmath$\fi}

\begin{titlepage}
\vspace*{-1cm}
\noindent
Revised ~August, 1994 \hfill{BGU-94 / 16 / June- PH}\\
\phantom{bla}
\hfill{UICHEP-TH/94-7}\\
\phantom{bla}
\hfill{hep-th/9407112}
\\
\begin{center}
{\large\bf Reduction of Toda Lattice  Hierarchy to \\
Generalized KdV Hierarchies and Two-Matrix Model}
\end{center}
\vskip .3in
\begin{center}
{ H. Aratyn\footnotemark
\footnotetext{Work supported in part by the U.S. Department of Energy
under contract DE-FG02-84ER40173}}
\par \vskip .1in \noindent
Department of Physics \\
University of Illinois at Chicago\\
845 W. Taylor St., Chicago, IL 60607-7059\\
{\em e-mail}: aratyn@uic.edu \\
\par \vskip .3in
{ E. Nissimov$^{\,2}$  and S. Pacheva \footnotemark
\footnotetext{On leave from: Institute of Nuclear Research and Nuclear
Energy, Boul. Tsarigradsko Chausee 72, BG-1784 $\;$Sofia,
Bulgaria. }}
\par \vskip .1in \noindent
Department of Physics, Ben-Gurion University of the Negev \\
Box 653, IL-84105 $\;$Beer Sheva, Israel \\
{\em e-mail}: emil@bguvms.bitnet, svetlana@bgumail.bgu.ac.il
\par \vskip .3in
A.H. Zimerman\footnotemark
\footnotetext{Work supported in part by CNPq}
\par \vskip .1in \noindent
Instituto de F\'{\i}sica Te\'{o}rica-UNESP\\
Rua Pamplona 145\\
01405-900 S\~{a}o Paulo, Brazil
\par \vskip .3in

\end{center}

\begin{abstract}

Toda lattice hierarchy and the associated matrix formulation of the
$2M$-boson KP hierarchies provide a framework for the Drinfeld-Sokolov
reduction scheme realized through Hamiltonian action within the second KP
Poisson bracket.
By working with free currents, which abelianize the second KP Hamiltonian
structure, we are able to obtain an unified formalism for the reduced
$SL(M+1,M-k)$-KdV hierarchies interpolating between the ordinary
KP and KdV hierarchies.
The corresponding Lax operators are given as superdeterminants
of graded $SL (M+1,M-k)$ matrices in the diagonal gauge
and we describe their bracket structure and field content. In particular,
we provide explicit free-field representations of the associated $W(M,M-k)$
Poisson bracket algebras generalizing the familiar nonlinear $W_{M+1}$-algebra.
Discrete B\"{a}cklund transformations for $SL(M+1,M-k)$-KdV are generated
naturally from lattice translations in the underlying Toda-like hierarchy.
As an application we demonstrate the equivalence of the two-matrix string
model to the $SL (M+1,1)$-KdV hierarchy.
\end{abstract}

\end{titlepage}

\noindent
\sect{Introduction}

Integrable Hamiltonian systems occupy an important place in diverse
branches of theoretical physics as exactly solvable models of
fundamental physical phenomena ranging from nonlinear hydrodynamics to
string theory of elementary particle's interactions at ultra-high energies
\ct{integr,Pol88,matrix}, including high-energy QCD in $D=3+1$ space-time
dimensions \ct{Fad-QCD}. Among the most notable
physically relevant integrable models is the Kadomtsev-Petviashvili (KP)
hierarchy of integrable soliton nonlinear evolution equations
\ct{Zakh,Dickey}. The main interest towards the KP hierarchy in the last few
years stems from their deep connection with string (multi-)matrix models
\ct{integr-matrix}.

The general KP system is an $1+1$-dimensional integrable model containing an
infinite number of fields. Its reductions with finite number of fields
(``multi-boson KP hierarchies'' for short) are, similarly, integrable
systems which naturally appear in two different but related settings.
As continuous (field-theoretic) integrable models they define a consistent
{\em Poisson reduction} of the complete continuous KP system \ct{ANP93}.
However, they can also be traced back to the {\em discrete} Toda
lattice hierarchy \ct{BX9204,BX9209,BX9212,BX9305}.
These two distinct formulations are in agreement due to the existence
of a discrete symmetry for the continuous multi-boson KP hierarchy,
which is a canonical mapping realized by a similarity transformation for the
underlying Lax operators \ct{discrete,similar}.
The presence of this discrete transformation allows us to view the Toda
lattice \ct{toda} as a union of sites, each being a gauge copy of
one continuous multi-boson KP system.
Throughout this paper we shall benefit essentially from this relation between
the Toda lattice and the continuous multi-boson KP system.

It has long been known that the most simple example of the multi-boson
KP systems, the two-boson KP system, contains the usual KdV
hierarchy \ct{BAK85}.
In \ct{AFGMZ} the Dirac reduction of the two-boson KP system was put up
and the KdV hierarchy was obtained in this process.
Subsequently, the Dirac reduction was applied to other multi-boson
KP systems which resulted in a large family of generalized KP-KdV
hierarchies \ct{BX9305,BX9311}.

In this paper we investigate the reduction of the multi-boson KP hierarchies
employing the Drinfeld-Sokolov (DS) reduction scheme realized
in a non-conventional way -- as Hamiltonian reduction within the
second KP Poisson bracket structure.
By working with free currents abelianizing the latter highly reducible
Poisson structure,
we are able to obtain an unified description of the various reduced
hierarchies, their bracket structure and their field content.
Let us recall that the KP hierarchy is endowed with bi-Hamiltonian
Poisson bracket structures resulting from the two compatible Hamiltonian
structures on the algebra of pseudo-differential operators \ct{STS83}.
The latter are given by:
\br
{\pbbr{\me{L}{X}}{\me{L}{Y}}}_1 \eq
- \llangle L \bv \left\lb X,\, Y \right\rb \rrangle  \lab{first-KP}\\
{\pbbr{\me{L}{X}}{\me{L}{Y}}}_2 \eq {\Tr}_A \( \( LX\)_{+} LY -
\( XL\)_{+} YL \)          \lab{second-KP-0}
\er
Here and below the following notations are used.
$<\cdot \v \cdot >$ denotes the standard bilinear pairing via
the Adler trace $\me{L}{X} = {\Tr}_A \( LX\)$ with
${\Tr}_A X = \int {\rm Res} X $. Here
$L,X,Y$ are arbitrary elements of the algebra of pseudo-differential
operators of the form $L = \sum_{k \geq -\infty} u_k D^k$,
$X = \sum_{k \geq - \infty} D^k X_k $, where
$D= \pa/ \pa x$ denotes the differential operator w.r.t. $x$.
Furthermore, the subscripts $\pm$ in $X_{\pm}\,$ denote taking the purely
differential or the purely pseudo-differential part of $X$, respectively.
The Lax operator of the KP hierarchy has the following specific form
{}~$L=D + \sum_{k=1}^{\infty} u_k D^{-k}$ , and therefore, the second Poisson
bracket \rf{second-KP-0} is modified to:
\br
{\pbbr{\me{L}{X}}{\me{L}{Y}}}_2 \eq {\Tr}_A \( \( LX\)_{+} LY -
\( XL\)_{+} YL \) \nonu  \\
&+& \!\!\int dx \, {\rm Res}\Bigl( \sbr{L}{X}\Bigr) \pa^{-1}
{\rm Res}\Bigl( \sbr{L}{Y}\Bigr) \qquad     \lab{second-KP}
\er
The last term is a Dirac-bracket term originating from the second-class
constraint $u_0 =0$ .

The Toda lattice hierarchy in the matrix form is a central object in
our study of Hamiltonian reduction.
Up to a phase-gauge transformation, the Toda matrix of the associated
linear problem has the form
of a matrix in the DS gauge with an extra traceless diagonal part.
Hence, there exists a residual gauge symmetry preserving the DS-like form of
the Toda matrix. Correspondingly, the space of Toda matrices splits into
orbits of this residual gauge group.
The reduction is then accomplished in the final step by restricting
to the symplectic quotient space (by quotienting out the residual
gauge symmetry).
In the case under consideration the final step involves removal of the
diagonal terms (currents) of the Toda hierarchy matrix. One obtains in this
process various generalized KP-KdV hierarchies with the usual KdV model
corresponding to the case when all the diagonal currents are gauged away.

Such a reduction, when based on matrix calculations,
becomes quickly cumbersome with the increasing rank of the matrices.
The question is whether there exists a convenient way of handling
the residual gauge transformations.
The natural symplectic Kirillov-Kostant-Symes (KKS) form associated with
the space of Toda matrices degenerates on the vector fields
tangent to the orbits defined by the residual gauge transformations.
Hence, one does not expect the action of the residual gauge transformation
to be Hamiltonian as there is no natural KKS-type Poisson bracket associated
with the linear Toda matrix problem before the last step of Hamiltonian
reduction is taken.
Surprisingly, it turns out that the relevant gauge group action is
nevertheless Hamiltonian but
with respect to the second Poisson bracket \rf{second-KP} of the
multi-boson KP system.
This enables us to describe the DS reduction within the framework
of the Poisson manifolds and to find closed expressions for the gauge
transformed quantities on the reduced manifold.

The abelianized representation, {\sl i.e.}, the representation in terms of
free currents, of the second Poisson bracket \ct{ANP9401} plays a key
r{\^{o}}le in the above formalism.
Here the underlying lattice structure is the Volterra lattice and the problem
is transformed from the DS gauge of the Toda hierarchy to the diagonal
gauge. In this representation the residual gauge
transformations take a simple form and the constraint manifold is described
directly in terms of the original abelian canonical variables.
It is crucial that the second bracket structure is reducible and
the residual gauge symmetry of DS problem triggers a total factorization of
the bracket structure.

The corresponding Lax operators appearing on various levels of reduction are
constructed in terms of currents spanning the Cartan subalgebra
of the graded $SL(M+1,M-k)$ Kac-Moody algebra \ct{fyu}.
The variable $k$ labels the level of reduction with $k=0$ corresponding to
the original $2M$-boson KP system and $k=M$ describing the maximal
reduction to the ordinary $SL(M+1)$-KdV hierarchy.
In the latter case the Lax operator reduces to the simple
determinant of the Fateev-Lukyanov type \ct{FL88}.
Hence, we obtain an unified formalism for the reduced
$SL(M+1,M-k)$-KdV hierarchies (KP-KdV hierarchies), which interpolate between
the original KP systems and the ordinary KdV hierarchies.
The generic Lax operators are given as superdeterminants
of the graded $SL (M+1,M-k)$ matrices in the diagonal gauge.
We describe their bracket structure.
Thanks to our abelianization technique we are also able to give a free-field
construction for all dynamical variables of the $SL(M+1,M-k)$-KdV hierarchies.

The generalized $SL(M+1,1)$-KdV hierarchies have recently been encountered
in the study of the two-matrix model \ct{BX9311a,enjoy}.
In \ct{enjoy} the simplest nontrivial Toda-like lattice integrable
system, derived from the partition function of the two-matrix model
with matrix potentials of orders $p_1 =$ arbitrary and $p_2 =3$,
was shown to be equivalent to the $1+1$ dimensional generalized $SL(3,1)$-KdV
hierarchy. In this paper we extend the above analysis to the case of
arbitrary finite $p_2$.

The organization of the material is as follows.
In Section 2 we recapitulate the basic facts
about the Toda lattice hierarchy and the matrix approach to the spectral
problem versus the continuum multi-boson KP hierarchy.
In Section 3 we compare Dirac and DS reductions of the two-boson
KP hierarchy and present DS reduction for the four-boson hierarchy.
Next, in Section 4 we show that the residual gauge transformation has a
Hamiltonian action with respect to the second KP Poisson bracket and discuss
how the DS reduction, described in the previous section, is induced in this
Hamiltonian manner.
These results are generalized to an arbitrary multi-boson KP hierarchy in
Section 5, where use is made of a set of free currents abelianizing the second
Poisson bracket structure.
These currents enter into the Lax operator in a form which naturally
leads to the graded $SL (M+1,M)$ Kac-Moody algebra.
In Section 6 the reduction process is shown to be equivalent to reducing
the graded $SL (M+1,M)$ algebra to $SL (M+1,M-k)$ algebra.
This framework allows for a simple expression for the second bracket
structure of the $SL(M+1,M-k)$-KdV hierarchy in terms of Lax operators
being superdeterminants of the graded $SL (M+1,M-k)$ matrices in the
diagonal gauge. Also, we show that the reduced generalized KP-KdV hierarchies
are integrable (bi-Hamiltonian) and possess canonical discrete symmetries.
Our construction provides explicit free-field representations of the
associated $W(M,M-k)$ Poisson bracket algebras generalizing the
well-known nonlinear $W_{M+1}$ algebra \ct{Zam}.
Discrete B\"{a}cklund transformations for $SL(M+1,M-k)$-KdV are generated
naturally from lattice translations in the underlying Toda-like hierarchy.
Finally, as an application we demonstrate in Section 7 the equivalence of
the two-matrix string model to the $SL (p_1 ,1)$-KdV hierarchy,
where $p_{1,2} \; \( p_1 \leq p_2\)$ are the orders of the matrix-model
potentials.

\sect{Toda Hierarchy versus Multi-Boson KP Hierarchy.}
\subsection{Toda Hierarchy and Matrix Approach to the Spectral Problem.}
We start with the spectral equation:
\br
\pa \Psi_n \eq \Psi_{n+1} + a_0 (n) \Psi_n     \lab{spectr}\\
\l \Psi_n \eq \Psi_{n+1}  + a_0 (n) \Psi_n
+ \sum_{k=1}^M a_k (n) \Psi_{n-k} \nonu
\er
associated with the Toda lattice hierarchy (for the most general case, see
ref.\ct{U-T}). Here $\pa \equiv \pa_x \equiv \pa/\pa t_{1,1}$, where
$t_{1,1}$ denotes the first lattice evolution parameter which is now
considered as a space coordinate of an $1+1$-dimensional integrable system.
The spectral equation \rf{spectr} can be rewritten as a matrix equation
$\( \one \pa - {\bf Q}\) \Psi =0$, which in components is given by:
\be
\left(\begin{array}{ccccc} \pa- a_0 (n-M) & -1 &0 & \ldots &0 \\
0& \pa- a_0 (n-M+1) & -1 &0 & \ldots \\
\vdots & \vdots &\ddots &\ddots &\vdots \\
0& 0&0  &\pa- a_0 (n-1)  & -1  \\
a_M (n) & a_{M-1} (n)& \ldots & a_1 (n) &\pa - \l \end{array} \right)
\left(\begin{array}{c}
\Psi_{n-M} \\ \Psi_{n-M+1} \\ \vdots \\ \Psi_{n-1} \\
\Psi_{n} \end{array} \right)  = 0
\lab{matrix}
\ee
Note first that by eliminating all $\Psi_{n-i}\;,\;i \ne 0$ from the set
of equations represented by \rf{matrix} we obtain:
\be
\l \Psi_n = L_n^{(M)} \Psi_{n}
\lab{spectrb}
\ee
where
\be
L_n^{(M)} = \pa + \sum_{k=1}^M a_k (n) {1 \over \pa - a_0 (n-k)}
\ldots {1 \over \pa - a_0 (n-1)}
\lab{lnn}
\ee
The matrix form \rf{matrix} of the Toda spectral problem will be
a starting point of our discussion of DS reduction.
We first perform a phase gauge transformation:
\be
\Psi_{n-k} \to \exp \( - {1 \over M+1} \sum_{i=1}^M \int a_0 (n-i)\)
\Psi_{n-k} \qquad,\qquad 0 \leq k \leq M
\lab{phase}
\ee
which transforms the diagonal terms in \rf{matrix} to:
\br
\pa- a_0 (n-i) &\to& \pa+{1 \over M+1} \sum_{j=1,j \ne i}^M a_0 (n-j)
-{M \over M+1} \; a_0 (n-i)  \lab{phasdia}\\
\pa- \l &\to& \pa- \l +{1 \over M+1} \sum_{j=1}^M a_0 (n-j) \nonu
\er
This transformation renders the matrix ${\bf Q}$ traceless (for $\l =0$).
Let us denote by
\be
\cM \Psi \equiv \( \pa - {\bf {\bar Q}}\) \Psi = 0 \quad , \quad
{\bf {\bar Q}} = \cE - \om \;\; ,\;\; \cE_{ij} = \d_{i+1,j} \;\; ,\;\;
\tr \om =0 \lab{DS-like}
\ee
the equation obtained from \rf{matrix}
by the above phase transformation accompanied by setting $ \l = 0$.
Now, one can follow the ideas of the DS Hamiltonian reduction scheme
\ct{DS,LOR}. Indeed, the space of Toda matrices
$\cO_{Toda}=\Bigl\{ {\bf {\bar Q}}\; ; {\bf {\bar Q}} ~{\rm as ~in
{}~\rf{DS-like}}\;\Bigr\}$ can be viewed as a submanifold of the phase space
of $SL(M+1)$ WZNW model, which is a coadjoint orbit of $SL(M+1)$ :
\be
\cO_G = \Bigl\{ S(g)=\pa g \, g^{-1} \; ; g \in G = SL(M+1) \Bigr\} \lab{orbit}
\ee
Thus, this submanifold $\cO_{Toda}$
corresponds to a {\em partial} gauge fixing of the first-class
constraints ~$\Phi \equiv \pa g\, g^{-1} \bgv_{\cG_{+}} - \cE =0 \; ,
\; g \in SL(M+1)$, in a gauged $SL(M+1)$ WZNW model whose ``big'' phase space
is \rf{orbit}. Here as usual $\cG_{\pm} \subset
\cG = sl (M+1)\,$ denote the nilpotent upper/lower-triangular
subalgebras. Therefore, there exists a residual $M$-dimensional
gauge symmetry group $\G \subset SL(M+1)\, ,\, h \in \G$ :
\be
\cM \equiv \pa - \cE + \om \to  h^{-1} \cM h
= \pa - \cE + {\bar \om}        \lab{resgau}
\ee
which preserves the form of the $\om$ matrix
and defines a gauge orbit for the Toda lattice hierarchy in the matrix form.
The natural symplectic form (KKS form) on the ``big'' phase space
\rf{orbit} degenerates on the vector fields tangent to the gauge orbits
\rf{resgau}. Hence, there is no natural KKS-type Poisson bracket structure
for the coefficients $a_0 (n-l), a_l (n)\; ,\; l=1,\ldots ,M\,$ of
${\bf {\bar Q}}$ associated with the linear matrix problem \rf{DS-like}.
Consequently, {\em a priori} one does not expect the action of the residual
gauge transformations \rf{resgau} to be Hamiltonian.
We shall find later on that this gauge action can nevertheless
be realized as a Hamiltonian action, namely, it is
generated by
the second KP Poisson bracket \rf{second-KP} for the $2M$-boson KP Lax
operator \rf{lnn} inherent to the Toda matrix spectral problem \rf{matrix}.

The determinant of $\cM$ in \rf{DS-like} can be written as:
\be
\det \cM = \pa^{M+1} + u_{M-1} \pa^{M-1} + \ldots + u_1 \pa +u_0
\lab{detm}
\ee
and clearly is invariant under \rf{resgau}. The differential operator
$\det \cM $ can also be obtained from $\cM \Psi = 0$ \rf{DS-like}
by eliminating all
$\Psi_{n-M+1}, \ldots, \Psi_{n}$ apart from $\Psi_{n-M}$.
Generally we obtain a family of Lax operators in the process of
eliminating all $\Psi$'s apart from one element of the column in
\rf{matrix}, which we denote by $\Psi_{n-i_0}$.
The case $i_0=M$ is the ``pure'' KdV Lax operator from \rf{detm} while $i_0=0$
gives the KP Lax operator $L_n^{(M)}$ from \rf{lnn} (up to a phase-gauge
transformation).
For $0<i_0<M$ we get a family of Lax operators invariant under various
subgroups of the residual gauge symmetry \rf{resgau}.
We shall implement in this paper the DS reduction scheme in the above
mentioned non-conventional setting -- as a Hamiltonian reduction with respect
to the second KP Poisson bracket, to describe this family of Lax operators
contained in the linear system of \rf{matrix}.

\subsection{KP Hierarchy: the First Bracket}
{}From \rf{spectr} we obtain the consistency conditions:
\br
\pa a_0 (n) \eq a_1 (n+1) - a_1 (n)  \lab{nt1eqa} \\           \
a_k (n) \eq a_{k} (n-1) + \biggl(\pa + a_0 (n-k) - a_0 (n-1) \biggr)
a_{k-1} (n-1) \quad \, k=1,\ldots,M
\lab{toda}\\
\pa a_M (n) \eq a_M (n)  \( a_0 (n) - a_0 (n-M) \)       \lab{nt1eqc}
\er
which are the Toda equations of motion.
It is easy to see that all $a_k(n)$, $k=0,1, \ldots,M $, at each lattice
site $n$, are expressed as functionals of only $2M$ independent
functions, which can be chosen to be, {\sl e.g.}: $a_0(M-k)$ and $a_k(M)$,
$k=1, \ldots,M$.

We shall now relate the Toda lattice equation \rf{toda}
to the recurrence relations for the $2M$-boson KP Lax operators derived
in \ct{ANP93} and used there to abelianize the first Poisson bracket
structure \rf{first-KP}.
Let us introduce the following correspondence (with $n=M$ in \rf{toda}):
\br
&&A^{(M)}_{M-k+1} \sim a_k (M) \quad;\quad
B^{(M)}_{M-k+1} \sim a_0 (M-k)
\qquad\;k = 1,\ldots,M
\lab{corres}\\
&&B^{(M-1)}_l \sim  a_0 (l-1) -a_0 (M-1) \qquad\;l = 1,\ldots,M-1
\lab{corres-a}
\er
Define now $a_M \equiv A^{(M)}_{M} \sim a_1(M)$
and $b_M \equiv B^{(M)}_{M}  \sim a_0 (M-1)$.
As a consequence of the last definition and \rf{corres-a} we find
the recurrence relation $B^{(M)}_{l} = b_M + B^{(M-1)}_l$.
Furthermore, we notice that identifications made in
\rf{corres}-\rf{corres-a} allow to recast the lattice Toda equation of
motion \rf{toda}  in the form of recurrence relations:
\be
\AM{l} = A_{l-1}^{(M-1)} + \( \pa + B_l^{(M-1)} \) A_l^{(M-1)}
\qquad  (l=2,\ldots ,M-1)  \lab{recur}
\ee
Assuming furthermore that $A_{0}^{(M)} = a_{M+1}(M)$
we get in addition:
\be
A_{1}^{(M)} = \( \pa + B_{1}^{(M-1)}\)  A_{1}^{(M-1)}
\lab{recur-a}
\ee

The above recurrence relations have been shown \ct{ANP93}
to be equivalent to the recursive formula for the $2M$-boson KP Lax operator
valid for arbitrary $M=1,2,3,\ldots$ (with $L_0 \equiv D\, ,\,
a_0 \equiv 0$) :
\br
L_{M} &\equiv& L_M (a,b) \equiv L_M \(a_1 ,b_1 ; \ldots
; a_M ,b_M \)  \nonu  \\
L_{M} \eq e^{\int b_{M}} \Bigl\lb b_{M} +
(a_{M} - a_{M-1} )D^{-1} + D L_{M-1} D^{-1} \Bigr\rb
e^{-\int b_{M}}                   \lab{3-1}
\er
In fact, the solution to \rf{3-1} is the $2M$-field Lax operator of the
form of \rf{lnn} \ct{BX9212}:
\be
L_{M} = D + \sum_{l=1}^{M} \AM{l}
\( D - \BM{l}\)^{-1} \( D - \BM{l+1}\)^{-1} \cdot\cdot\cdot
\( D - \BM{M}\)^{-1}    \lab{3-3}
\ee
with coefficients satisfying \rf{recur} and \rf{recur-a}. As a result,
the latter are
expressed in terms of the free boson fields $\, \( a_r , b_r \)_{r=1}^M \,$
spanning Heisenberg Poisson bracket algebra :
\be
\lcurl a_r (x),\, b_s (y) \rcurl_{P^{\pr}} =
 - \d_{rs} \pa_x \d (x-y)          \lab{3-2}
\ee
as
\br
\BM{l} \eq \sum_{s=l}^{M} b_s \qquad \qquad , \qquad \qquad
\AM{M} = a_{M}
\lab{3-4}  \\
\AM{M-r} \eq \sum_{n_r =r}^{M-1} \cdot\cdot\cdot \sum_{n_2 =2}^{n_3 -1}
\sum_{n_1 =1}^{n_2 -1} \( \pa + b_{n_r} + \cdot\cdot\cdot +
b_{n_r -r +1} \) \cdot\cdot\cdot \( \pa + b_{n_2} + b_{n_2 -1}\)
\( \pa + b_{n_1} \) a_{n_1} \nonu 
\er

This recursive construction of the Lax in \rf{3-3} leads
to the following \ct{ANP93}:
\lskip
\prop {\em The $2M$-field Lax operators
\rf{3-3} are consistent Poisson reductions
of the full KP Lax operator
for any $\, M=1,2,3,\ldots \,$} .

Thus, the first Poisson bracket structure for $L_{M}$ from \rf{3-3} is given
by:
\be
\Bigl\{ \llangle L_{M} \bv X \rrangle \, ,\,
\llangle L_{M} \bv Y \rrangle \Bigr\}_{P^{\pr}} =
- \llangle L_{M} \bv \left\lb X,\, Y
\right\rb \rrangle    \lab{3-6}
\ee
where $\, X,\, Y\,$ are arbitrary fixed elements of the algebra
of pseudo-differential operators.
The subscript $P^{\pr}$ in \rf{3-6} indicates that the constituents of
$L_M (a,b)$ satisfy \rf{3-2}.

\sect{Reductions of the Two-Boson KP Hierarchy to KdV}
\subsection{Two-boson KP Hierarchy and the Dirac Reduction}
We shall consider here truncated elements of the KP hierarchy providing the
simplest example of \rf{3-3} and given by Lax operator of the form
\be
L_1 = D + a \(D - b \)^{-1}  \lab{L-ab}
\ee
with two free Bose currents $(a,b )$ \ct{BAK85,2boson}.
The Lax operator can be cast into the standard form
$L_1 =D + \sumi{n=0} w_n D^{-1-n}$
with coefficients $w_n = (-1)^n a (D - b)^n \cdot 1$.
A calculation of the Poisson bracket structures using definition
\rf{first-KP} and \rf{3-6}
yields the first bracket structure  of two-boson $(a,b)$ system:
$\pbr{a(x)}{b(y)}_1= - \d^{\pr} (x-y)$, and zero otherwise.
This implies that the coefficients $w_n (a,b)$ of $L_1$, as functionals of
$a,b$, satisfy the linear $\Win1\,$ Poisson-bracket algebra.
The second bracket structure \rf{second-KP} takes in this case the form:
\br
\{ a (x) \, , \, b (y) \}_2 &=& -b(x) \d^{\pr} (x-y) - \d^{\pr\pr} (x-y)
\nonu\\
\{ a (x) \, , \, a (y) \}_2 &= &  -2 a (x) \d^{\pr} (x-y) -a^{\pr}
(x) \d (x-y) \lab{2pab}\\
\{ b (x) \, , \, b (y) \}_2 &=&- 2\, \d^{\pr} (x-y) \nonu
\er
Now, based on this bracket, $w_n (a,b)$ satisfy the nonlinear $\hWinf\,$
Poisson-bracket algebra.

In \ct{AFGMZ,talk} we applied the Dirac reduction scheme to obtain one-boson
KdV hierarchy from the two-boson KP hierarchy.
The general feature is a transformation
of the two-boson Hamiltonian equations of motion expressed
in terms of the 2-nd bracket
structure $\d Z/ \d {t_r} = \{ Z \,, \, H_r \}_2$
(where $Z$ denotes the original degrees of freedom)
to one-boson Hamiltonian system according to the Dirac scheme:
\be
\partder{X}{t_r} = \{ X \,, \, H^{D}_r \}_{Dirac}
\lab{diraham}
\ee
with $X$ denoting a surviving one-boson degree of freedom.

Consider the Dirac constraint: $\Theta \equiv b=0$ for the system described by
$L_1$. First, let us discuss the resulting Dirac bracket structure.
We find for the surviving variable $a$:
\br
\{ a (x) \, , \, a (y) \}_2^{D} \eq \{ a (x) \, , \, a (y) \}_2
- \int dz dz^{\pr} \{ a (x) \, , \, \Theta (z) \}_2 \{ \Theta (z) , \Theta
 (z^{\pr} )
\}_2^{-1} \{ \Theta (z^{\pr}) \, , \, a (y) \}_2  \nonu \\
\eq -\( 2 a (x) \pa +a^{\pr} (x) + \h \pa^3\) \d (x-y)
\lab{dira2}
\er
The reduced Lax operator looks now as:
\be
l = D + a D^{-1}  \lab{faa3}
\ee
and the corresponding (non-zero) lowest Hamiltonian functions $H^{KdV}_r
\equiv \Tr l^r /r$ are
\be
H^{KdV}_{1} = \int a \quad ;\quad H^{KdV}_{3} = \int a^2
\quad ;\quad H^{KdV}_{5} = \int \( 2 a^3 + a a^{\pr \pr} \)
\lab{kdvham}
\ee
Moreover one checks that the flow equation:
\be
{\d l}/ {\d t_r} = \sbr{(l^r)_{+} }{l}
\lab{flowkdv}
\ee
gives on the lowest level
${\d a}/ {\d t_1}= a^{\pr}$ and ${\d a}/ {\d t_3} =
a^{\pr \pr \pr} + 6 a a^{\pr}$,
where the last equation is the well-known KdV equation.

We shall now demonstrate that the DS reduction is an alternative to the
Dirac reduction of the  two-boson KP hierarchy to the usual KdV hierarchy
\ct{talk}.

\subsection{Matrix form of Two-Boson KP Hierarchy and the DS Reduction}
One can associate $sl(2)$ matrices to pseudo-differential Lax operators
in the following way \ct{talk} :
\be
L = D + A +B \, D^{-1} \, C \; \; \sim \; \;
{\cal A} = \fourmat{-\h A}{-C}{B}{\h A}
\lab{laxsl2}
\ee
so that the gauge transformation of the Lax operator
$L^{\pr}  = e^{-\chi}L e^{\chi}$ corresponds to $SL(2)$ gauge transformation
${\cal A}^{\pr} = g {\cal A} g^{-1} + g \pa g^{-1}$
with a diagonal $2\times2$-real
unimodular matrix $ g = {\rm diag} ( \exp {\chi/2}, \exp -{\chi/2} )$.

To see the connection with the matrix Toda hierarchy (with $\l=0$):
\be
\fourmat{\pa - a_0 (n-1)}{-1}{a_1 (n)}{\pa} \twocol{\Psi_{n-1}}{\Psi_{n}}
= 0
\lab{mattoda}
\ee
let us introduce new variables as in \rf{phase}:
\be
\twocol{e^{\h \int a_0 (n-1)} \Psi_{n-1}}{e^{\h \int a_0 (n-1)}\Psi_{n}}
\lab{newvar}
\ee
and denote $a_0 (n-1)=b,a_1 (n)= a$. According to \rf{laxsl2},
we find the association:
\be
L_{KP} = D + b + a D^{-1} \, \sim \,
\cA_{KP} = \fourmat{-\h b}{-1}{a}{\h b}   \lab{KPsl2}
\ee
Here, the important point is that there is a residual gauge transformation
generated by:
\be
g_0 \equiv \fourmat{1}{0}{\g}{1}
\lab{resi}
\ee
which preserves the form of $\cA_{KP}$ under:
\be
{\cal A}^{\pr} = g_0^{-1} {\cal A} g_0 + g_0^{-1} \pa g_0 =
\fourmat{-\h b-\g}{-1}{a+ \g b + \g^2 + \g^{\pr} }{\h b+\g}
\lab{KPsl2pr}
\ee
Let us analyze what is happening by using the
DS formalism. Consider the space of first-order differential operators
with coefficients being $2\times 2$ matrices:
\be
M_{\cE} = \lcurl \cD^{(1)} = D - \cE + \om \bv \cE = \fourmat{0}{1}{0}{0}
\; , \; \om =  \fourmat{\om_{11}}{0}{\om_{21}}{\om_{22}}
\rcurl
\lab{dsm}
\ee
and the group
\be
\G \equiv \lcurl \G \bv \G \equiv \fourmat{1}{0}{\g}{1} \rcurl
\lab{resid}
\ee
acting on $M_{\cE}$ according to
\be
\G^{-1} \( D - \cE + \om \) \G = D - \cE + \bom
\lab{dceom}
\ee
with
\be
\bom =  \fourmat{\om_{11}-\g}{0}{\om_{21} + \g (\om_{22} - \om_{11})
+ \g^2 + \g^{\pr}}{\om_{22}+\g}
\lab{ompr}
\ee
In the spirit of Hamiltonian reduction consider the
quotient space $M_{\rm red} = M_{\cE} / \G$.
There exists a convenient realization of $M_{\rm red}$ in terms of
second-order differential operators with scalar coefficients.
The procedure to obtain it goes as follows. Consider the equation:
\be
\cD^{(1)} { \psi_1 \choose \psi_2} =  0
\lab{dspsi}
\ee
Eliminating $\psi_2$ from this equation we arrive at $L^{(2)} \psi_1=0$ with
\be
L^{(2)} \equiv \det \( \cD^{(1)}\) = D^2 + ( \om_{11} + \om_{22} ) D
+ \om_{21} + \om_{11} \om_{22} + \om_{11}^{\pr}
\lab{l2}
\ee
Clearly $\det \( \G^{-1} \cD^{(1)} \G\) = \det \( \cD^{(1)}\)$ and,
therefore, the space of the second order differential operators of the
form \rf{l2} parametrizes the quotient space $M_{\rm red} $.

Let us study now the special case of two-boson $sl(2)$ matrix:
\be
\om = \fourmat{\om_{11}}{0}{\om_{21}}{\om_{22}} =
\fourmat{-\h b}{0}{a}{\h b}
\lab{omj}
\ee
For $\G$ with $\g= -\h b$ the transformed $\bom$ matrix
\be
\bom = \fourmat{0}{0}{u }{0}
= \fourmat{0}{0}{a- {1\over 4} b^2 - \h b^{\pr} }{0}
\lab{diagom}
\ee
has diagonal elements equal to zero. It means that, according to \rf{KPsl2},
the associated Lax operator is:
\be
L_{KP} = D + u D^{-1} \qquad {\rm with} \qquad
u = a- {1\over 4} b^2 - \h b^{\pr}
\lab{udef}
\ee
One can check that, with $(a , b)$ satisfying the second Poisson bracket
\rf{2pab}, $u$  commutes with $b$ and satisfies the Virasoro algebra:
\be
\pbr{u(x)}{u(y)} = -2 u (x)\, \d^{\pr} (x-y) - u^{\pr} (x) \,\d (x-y) -
\h \d^{\pr \pr \pr} (x-y)
\lab{uvira}
\ee
We also note that with $\om$ like in \rf{omj} the second-order
differential operator \rf{l2} becomes a typical KdV operator $L^{(2)} =
D^2 + u$.
Hence, the first-order DS operator $\cD^{(1)}$ \rf{dsm} with $\bom$
as in \rf{diagom}, or its associated Lax operator $L_{KP}$
{}from \rf{udef}, represent just a special gauge choice on $M_{\cE}$
equivalent to the KdV Lax operator $L^{(2)}$.

\subsection{Drinfeld-Sokolov Reductions of Four-Boson KP Hierarchy}
Start again with the Toda matrix problem with $\l=0$
\be
\ttmat{\pa - a_0 (n-2)}{-1}{0}{0}{\pa - a_0 (n-1)}{-1}{a_2 (n)}{a_1 (n)}{\pa}
\thrcol{\Psi_{n-2}}{\Psi_{n-1}}{\Psi_{n}} =\, 0
\lab{todasixmat}
\ee

Consider a general space of first order differential operators
with coefficients being $3\times 3$ matrices:
\be
M_{\cE} = \lcurl \cD^{(2)} \equiv D - \cE + \om \rcurl
\lab{dsm3}
\ee
with
\be
\cE = \ttmat{0}{1}{0}{0}{0}{1}{0}{0}{0}
\lab{ce3}
\ee
and
\be
\om =  \ttmat{J_1}{0}{0}{0}{J_2}{0}{A_1}{A_2}{J_3}
\lab{om3}
\ee
Eliminating $\psi_1,\psi_2$ from the equation:
\be
\cD^{(2)} \thrcol{\psi_1}{\psi_2}{\psi_3} = 0
\lab{d2psi}
\ee
we get the Lax operator of the four-boson KP hierarchy:
\be
\(\pa + A_2 {1\over \pa + J_2-J_3}+A_1 {1\over \pa +J_1-J_3}
{1\over \pa -J_2 -J_3}\) e^{ - \int J_3} \psi_3 = 0
\lab{lkp2psi}
\ee
while eliminating $\psi_2,\psi_3$ we get a KdV-type Lax operator:
\br
&&\( \pa^3 + u_0 \pa^2 + u_1 \pa +u_2 \) \psi_1 =0 \lab{kdv3}\\
&& u_0 \equiv J_1 +J_2 + J_3 \nonu\\
&& u_1 \equiv A_2 +2 J_1^{\pr} +J_2^{\pr} + J_1 J_2 +J_1 J_3 +J_2J_3 \nonu\\
&& u_2 \equiv A_1 + A_2 J_1 + J_1^{\pr\pr} +(J_1J_2)^{\pr} + J_1^{\pr} J_3
+ J_1 J_2 J_3\nonu
\er
Applying similar arguments as in subsection 3.2, we are led to consider a group
consisting of lower triangular $3 \times 3$ matrices
\be
\G \equiv \lcurl \G \bv \G \equiv \ttmat{1}{0}{0}{\a_1}{1}{0}{\a_2}{\a_3}{1}
 \rcurl
\lab{gamma3}
\ee
with an inverse
\be
\G^{-1} \equiv \ttmat{1}{0}{0}{-\a_1}{1}{0}{-\a_2+\a_3\a_1}{-\a_3}{1}
\lab{gaminv}
\ee
The action of $\G$ on $M_{\cE}$ according to \rf{dceom} produces the
following transformation of $\om$ \rf{om3} :
\be
\om \longrightarrow \bom=
\ttmat{J_1-\a_1}{0}{0}{\bom_{21}}{J_2+\a_1-\a_3}{0}
{\bom_{31}}{\bom_{32}}{J_3+\a_3}
\lab{bom}
\ee
where
\br
\bom_{21} \eq \a_1^{\pr} +\a_1^{2} - \a_2 + \a_1 (J_2-J_1) \lab{boms}\\
\bom_{31} \eq A_1 + A_2 \a_1 + \a_2^{\pr}
+ \a_3 \a_1 (J_1 -J_2) + \a_2 (\a_1+\a_3+J_3 -J_1) - \a_3( \a_1^2+\a_1^{\pr})
\nonu \\
\bom_{32} \eq A_2 + \a_3^{\pr} + \a_3 (J_3 - J_2 - \a_1) + \a_2 + \a_3^2
\nonu
\er
$\G$ will define the little group preserving the form of
$\cD^{(2)}$ if we impose $\bom_{21}=0$, {\sl i.e.},
\be
\a_2 = \a_1^{\pr} +\a_1^{2}  + \a_1 (J_2-J_1)
\lab{alpha2}
\ee
Hence the little group has only two independent components $\a_1,\a_3$
and the transformed matrix $\bom $ takes the form:
\be
\bom=
\ttmat{J_1-\a_1}{0}{0}{0}{J_2+\a_1-\a_3}{0}
{{\wti A}_1}{{\wti A}_2}{J_3+\a_3}
\lab{bom2}
\ee
with
\br
{\wti A}_2 \eq A_2 + \a_1^{\pr} + \a_3^{\pr} +\a_1^2+\a_3^2+ \a_1 (J_2 -J_1-
\a_3) +\a_3 (J_3 - J_2 )
\lab{baboms} \\
{\wti A}_1 \eq A_1 + \a_1^{\pr\pr} + \(\a_1^{\pr} +\a_1^2 \) (J_2+J_3 -2 J_1)
\nonu\\
&+&\a_1 \left\lbrack A_2 + \a_1^{2}+  3 \a_1^{\pr} +
 J_2^{\pr} - J_1^{\pr} + (J_2- J_1)(J_3-J_1)\, \right\rbrack
\nonu
\er

Let us return to the Toda problem \rf{todasixmat} whose matrix differential
operator belongs to the space $M_{\cE}$ \rf{dsm3}.
In order to agree with future convention let us introduce the
notations:
\be
a_0 (n-1) \equiv B_2 \;\; ,\;\; a_0 (n-2) \equiv B_1 \;\; ,\;\;
a_1 (n) \equiv A_2 \;\; ,\;\; a_2 (n) \equiv A_1 \lab{B-notat}
\ee
We can again gauge away the trace of the matrix appearing in the spectral
problem by letting $ \Psi \to \exp \(- \int (B_2 +B_1)/3 \) \Psi$.
This results in a differential operator $\cD^{(2)}$ \rf{dsm3} with
\underline{traceless} matrix $\om$ \rf{om3} where:
\be
J_1 = -(2B_1-B_2)/3 \quad;\quad J_2 = -(2B_2-B_1)/3\quad;\quad
J_3 = (B_2+B_1)/3
\lab{bes}
\ee
We can choose the two independent parameters $\a_1,\a_3$ in $\G$ \rf{gamma3}
to eliminate diagonal elements in \rf{bom} by taking:
\be
\a_1 = - {1 \over 3 } (2B_1-B_2) \qquad ;\qquad \a_3 =
- {1 \over 3 } (B_2+B_1)  \lab{choa}
\ee
In this case we arrive at the standard DS-gauge for $\cD^{(2)}$ :
\be
{\bar \cD}^{(2)}=\, \G^{-1} \( D - \cE + \om \) \G =
\ttmat{D}{-1}{0}{0}{D}{-1}{{\wti A}_1}{{\wti A}_2}{D} \quad;\quad
\det  {\bar \cD}^{(2)} = \pa^3 + {\wti A}_2 \pa + {\wti A}_1
\lab{bussima}
\ee
with
\br
{\wti A}_2 \eq A_2 - B_1^{\pr} - {1 \over 3 }\( B_2^2 + B_1^2 - B_2 B_1 \)
\lab{bussia1} \\
{\wti A}_1 \eq A_1 - {1 \over 3 } ( 2 B_1^{\pr\pr} -
B_2^{\pr\pr}) - {1 \over 3 } ( 2 B_1 - B_2)\( A_2 +B_1^{\pr} -B_2^{\pr}
+(B_1-B_2)B_1 \)
+ {2 \over 27} ( 2 B_1 - B_2)^3
\nonu
\er
{}From the second Poisson-bracket algebra of the four-boson KP hierarchy,
satisfied by $A_{1,2},B_{1,2}$,
we find that ${\wti A}_2, {\wti A}_1$ satisfy the Poisson
brackets of Bussinesq hierarchy (the $W_3$ algebra) (see e.g.
\ct{BX9305})
\br
\{{\wti A}_2(x), {\wti A}_2(y)\}_{2} \eq-(2 {\wti A}_2\pa +{\wti A}^{\pr}_2+
2\pa^3)\d (x-y) \nonu \\
\{{\wti A}_2(x), {\wti A}_1(y)\}_{2}\eq  - \(3{\wti A}_1\pa +2{\wti A}^{\pr}_1
-\pa^2 {\wti A}_2-\pa^4\)\d(x-y) \lab{w3alg}\\
\{{\wti A}_1(x), {\wti A}_1(y)\}_{2} \eq- \( 2{\wti A}^{\pr}_1\pa +
{\wti A}^{\pr\pr}_1-\frac{2}{3}({\wti A}_2+\pa^2)
(\pa {\wti A}_2+\pa^3)\) \d (x-y) \nonu
\er
The above reduction was complete in the sense that all the diagonal elements
were removed. In general, there are plenty of possibilities for a partial
reduction with, {\sl e.g.}, one diagonal element surviving the reduction.
We now describe one such a choice.
Note first that in the Toda hierarchy problem with a general $\om$ \rf{om3}
the diagonal terms $J_i$ can be represented as:
\be
J_1 = -B_1 + {\bar B} \quad;\quad
J_2= -B_2 + {\bar B} \quad;\quad
J_3= {\bar B} \lab{bbb}
\ee
where the term ${\bar B}$ indicates an addition which can be generated by
applying a gauge transformation (which effectively changes the trace of the
underlying matrix).

The particular choice of $\a_1, \a_3,{\bar B}$ we now make is to obtain
in \rf{bom2} :
\be
J_1-\a_1 = \h ( B_2 + B_1 )\quad;\quad
J_2 + \a_1 -\a_3= 0\quad;\quad
J_3+ \a_3= 0 \lab{bbb1}
\ee
The solution is $\a_3=- {\bar B}= -( B_2 + B_1 )/2$ and $\a_1 =- B_1$.
Inserting these parameters into \rf{baboms} we get by partial DS
reduction a hierarchy with three fields ${\wti A}_2 , {\wti A}_1, \cB =
- ( B_2 + B_1 )/2$ where
\br
{\wti A}_2 \eq A_2 - \h \(B_2^{\pr} +3 B_1^{\pr} \) - {1 \over 4} ( B_2
- B_1 )^2
\lab{a1part} \\
{\wti A}_1 \eq A_1 - B_1^{\pr\pr}+ (B_1 B_2 )^{\pr} - B_1 A_2
\nonu
\er
We shall provide below (see \rf{a2alg}) the closed Poisson-bracket algebra
satisfied by these fields reproducing result obtained in \ct{BX9305} by a
Dirac method.

\sect{Drinfeld-Sokolov Reduction Induced by the KP Poisson Structure}
It turns out that the action of the residual gauge symmetry discussed above
in the  case of two- and four-boson KP hierarchies is Hamiltonian
(generated by a Poisson bracket structure). The relevant
bracket structure turns out to be the second KP Poisson structure
\rf{second-KP}. Before presenting the general framework we start with examples
of two- and four-boson KP hierarchies to reproduce results of the DS
reduction.

\subsection{The Case of Two-Bosons}
Recall the second bracket structure for two-boson system described by
\rf{2pab} and consider the abelian group generated by
\be
G \, = \, \exp \(- \int b\; \b \)
\lab{g2boson}
\ee
which acts on $( a, b)$ through the second bracket, {\sl e.g.} :
\be
G^{-1} b (x) G =  b (x) - \{b(x), \int b \b \, \}_2
+ \h \{ \{b(x), \int b \b \, \}_2\, , \, \int b \b \, \}_2 + \ldots
\lab{gjg1}
\ee
A simple calculation shows that the result of such an action on $( a, b)$ is:
\be
b \to G^{-1} b (x) G = b + 2 \b^{\pr} \quad;\quad
a \to G^{-1} a (x) G = a + b \b^{\pr} + \b^{\pr\pr} + (\b^{\pr})^2
\lab{jjbeta}
\ee
Choosing $\b^{\pr}= \g$ we see that \rf{jjbeta} reproduces the result
in \rf{KPsl2pr} as we set up to show.
Since this transformation is generated by the current $b(x)$ satisfying
Heisenberg Poisson-bracket algebra (last eq.\rf{2pab}),
we see that transforming
$b\to 0$ by the above transformation amounts to imposition of the Dirac
constraint $b =0$.
This explains why the above DS gauge choice in \rf{diagom}
leading to KdV was equivalent to taking the Dirac bracket.
This explanation is based on the curious equivalence between
transformations generated by lower triangular matrices acting on matrices
representing KP hierarchy and transformations generated
by currents acting on the second bracket Poisson manifold of the multi-boson KP
hierarchy.

Note, that the quantity $u = a- {1\over 4} b^2 - \h b^{\pr}$ parametrizing the
reduced manifold is invariant under the transformation \rf{jjbeta}
since $\pbr{u}{b}=0$, the point which is obvious in matrix
formulation but which can only be verified a posteriori for the current
generated transformations.
This fact is crucial in explaining why from algebraic point of view
the Dirac conditions $b=0, a =u$ agree with the gauge fixing as done above.
We shall see that this relationship is of general nature and applies to
arbitrary multi-boson KP systems.

\subsection{The Case of Four-Bosons}
In the case of four-boson KP hierarchy consider a group action on the KP
Poisson manifold generated by:
\be
G = \exp- \int  \, \( B_2 (\g_3 - \g_1) + B_1 \g_1 \)
\lab{fourg}
\ee
Our choice of the $\g$-parameters is dictated by the conventions of the
corresponding DS reduction as shown above.
The action of $G$ gives ({\sl e.g.}, $G^{-1} B_2 G$ etc.; cf. \rf{gjg1}) :
\br
B_2 &\to& B_2 + 2 \g_3^{\pr} - \g_1^{\pr} \nonu \\
B_1 &\to& B_1 + \g_3^{\pr} + \g_1^{\pr} \nonu \\
A_2 &\to& A_2 + \g_3^{\pr\pr} + \g_1^{\pr\pr} + B_2 (\g_3^{\pr} - \g_1^{\pr})
+ B_1 \g_1^{\pr} + (\g_3^{\pr})^2 + (\g_1^{\pr})^2  - \g_1^{\pr}\g_3^{\pr}
\lab{fbotransf}\\
A_1 &\to& A_1 + A_2 \g_1^{\pr} + \g_1^{\pr\pr\pr} + (2 B_1- B_2)
\g_1^{\pr\pr} +(B_1^{\pr}- B_2^{\pr}) \g_1^{\pr} \nonu \\
&+ &B_1 ( B_1- B_2) \g_1^{\pr} +3 \g_1^{\pr}\g_1^{\pr\pr}+
( 2 B_1 - B_2) \g_1^{\pr} \g_1^{\pr} + (\g_1^{\pr})^3
\nonu
\er
Taking $\g_i^{\pr} = \a_i$ we recover \rf{baboms} with the choice of
$B_i$'s as in \rf{bes}.
We can now eliminate $B_2$ and  $B_1$ by appropriate choice of $\a$'s
recovering the Boussinesq hierarchy of \rf{bussia1}.

Now, let us discuss the problem of invariance under $\a_1,\a_3$ transformations
on the reduced manifold.
First, we notice that ${\wti A}_1,{\wti A}_2$ of the Bussinesq hierarchy
\rf{bussia1}
are invariant under transformations given in \rf{fbotransf}.
The situation with partial gauge fixing is more subtle.
We shall examine the invariance under separate $\a_1$ and
$\a_3$ transformations by using the first two relations from \rf{fbotransf}.
There are two possibilities:
\begin{enumerate}
\item
$B_2 +B_1$ - invariant under $\a_1$-transformation
\item
$2 B_1 -B_2 $ - invariant under $\a_3$-transformation
\end{enumerate}
In case 1) we get as before three fields ${\wti A}_1 , {\wti A}_2, \cB =
- ( B_2 + B_1 )/2$, where $A_1 , A_2$ are given by \rf{a1part} --
both being invariant under $\a_1$-transformation.
This can be achieved by choosing (for instance)
$\a_3=-( B_2 + B_1 )/2$ and $\a_1 =- B_1$ so
$B_2 \to 0$ and $B_1 \to -( B_2 + B_1 )/2$.

In the second case we take $\a_1=0, \a_3=-\h B_2$ in order to have
transformations $B_2 \to 0$ and $B_1 \to \cB\equiv \h (2 B_1 -B_2) $.
With this choice we get:
\br
{\wti A}_2 \eq A_2 - \h B_2^{\pr} - {1 \over 4} B_2^2
\lab{a2part} \\
{\wti A}_1 \eq A_1
\nonu
\er
which could be obtained by putting $B_1=0$ in \rf{a1part} .
It is easy to see that \rf{a2part} is invariant under $\a_3$-transformation.
One easily verifies, that for $A_{1,2}, B_{1,2}$ satisfying the Poisson
brackets
dictated by \rf{second-KP}, the fields \rf{a2part} together with $\cB$
satisfy a closed algebra:
\br
\{{\wti A}_2(x), {\wti A}_2(y)\}_{2}\eq- (2{\wti A}_2\pa +{\wti A}^{\pr}_2
+\h\pa^3)\d (x-y) \nonu\\
\{{\wti A}_2(x), {\wti A}_1(y)\}_{2} \eq -(3{\wti A}_1\pa +
2{\wti A}^{\pr}_1)\d(x-y) \nonu\\
\{{\wti A}_2(x), \cB (y)\}_{2} \eq - (\frac{3}{2}\pa^2+\cB \pa)\d(x-y)
\lab{a2alg}\\
\{{\wti A}_1(x), {\wti A}_1(y)\}_{2}\eq -\lb (2{\wti A}^{\pr}_1+
4 {\wti A}_1 \cB )\pa
+{\wti A}^{\pr\pr}_1 + 2({\wti A}_1 \cB)^{\pr} \rb \d(x-y) \nonu \\
\{{\wti A}_1(x), \cB (y)\}_{2} \eq - \({\wti A}_2\pa +(\pa+ \cB)^2\pa \)
\d (x-y) \nonu\\
\{ \cB (x), \cB (y)\}_{2} \eq - \frac{3}{2}\d^{\pr} (x-y) \nonu
\er
obtained first by the Dirac bracket method in \ct{BX9305}.

\sect{Generalized Miura Transformation for Multi-Boson KP Hierarchies.}
The two-boson hierarchy given in Sect. 3 and described by $L_1$ \rf{L-ab}
is equivalent
to the model based on the pseudo-differential operator \ct{YW9111}:
\be
L_1 = \( D- e \)\( D- c\)\( D- e-c \)^{-1}
= D + \( e^{\pr} + ec \) \( D- e-c \)^{-1}
\lab{celax}
\ee
The Miura-like connection between these hierarchies
generalizes the usual Miura transformation
between one-bose KdV and mKdV structures and takes the form \ct{AFGMZ}:
\be
a = e^{\pr} + ec \qquad ; \qquad b= e+c
\lab{gmiura}
\ee
This Miura transformation $(e,c) \longrightarrow (a,b)$
can easily be seen to abelianize the second bracket \rf{2pab}, meaning that
whenever
\be
\pbr{e\,(x)}{c\,(y)}_2 \, =\,- \d^{\pr} (x-y)
\lab{p2ec}
\ee
then $a,b$, given by \rf{gmiura}, satisfy \rf{2pab}.

The above structures naturally appear in connection with the Toda and Volterra
lattice hierarchies \ct{toda}. Consider namely the spectral equation:
\be
\pa \Psi_n = \Psi_{n+1} + a_0 (n) \Psi_n   \quad;\quad
\l \Psi_n =\Psi_{n+1}  + a_0 (n) \Psi_n
+ a_1 (n) \Psi_{n-1}
\lab{psia} 
\ee
The Miura transformed hierarchy defined by \rf{celax} can be associated
with a ``square-root'' lattice with respect to the original Toda lattice
system \rf{psia}:
\br
\l^{1/2}\; {\wti \Psi}_{n+\h} = \Psi_{n+1} + \cA_{n+1} \Psi_n
\quad;\quad
\l^{1/2} \; \Psi_n = {\wti \Psi}_{n+\h} + \cB_n {\wti \Psi}_{n-\h}
\lab{sqra}   \\
{\wti \Psi}_{n+\h} = \( \pa - \cB_n - \cA_{n} \){\wti \Psi}_{n-\h}
\quad ;\quad   \Psi_{n+1} = \( \pa - \cB_n - \cA_{n+1} \) \Psi_{n}   \nonu
\er
which yields the Volterra chain equations \ct{toda}. Excluding the
half-integer modes in \rf{sqra} we recover \rf{psia} with:
\be
a_0 (n) = \cA_{n+1} + \cB_n \quad ,\quad
a_1 (n) = \cA_n \cB_n = \cB_{n-1} + \pa \cA_n   \lab{volt-toda}
\ee
where in the last eq.\rf{volt-toda} we have used one of the Volterra
equations of motion (consistency condition for the Volterra spectral problem
\rf{sqra}) :
\be
\pa \cA_n = \cA_n (\cB_n - \cB_{n-1} ) \lab{volt-eq}
\ee
Eqs.\rf{sqra} can be cast into the form:
\be
\l \Psi_n = L_n^{(1)} \Psi_{n} \quad ,\quad
L_n^{(1)} =
\( \pa - \cA_n \) \( \pa - \cB_{n-1} \) \( \pa - \cB_{n-1} - \cA_{n} \)^{-1}
\lab{ncelax}
\ee
which, upon the identification $\cA_n = e, \cB_{n-1}= c$, agrees with
\rf{celax}. Moreover, recalling the identification $a = a_1 (n)\; ,\;
b=a_0 (n-1)$ for the coefficients of $L_1$ \rf{L-ab} (cf. \rf{lnn}),
we see that the relation \rf{volt-toda} between the coefficients
of the Toda and Volterra discrete spectral problems precisely matches the
Miura relation \rf{gmiura} for the coefficients of the two-boson KP Lax
operators \rf{L-ab} and \rf{celax} abelianizing the first and the second KP
Poisson bracket structures, respectively.

Furthermore, using again the Volterra equation \rf{volt-eq} we can rewrite
\rf{ncelax} as:
\be
L^{(1)}_n = \( \pa - \cA_n \) \( \pa - \cB_{n} - \cA_{n} \)^{-1}
\( \pa - \cB_{n} \) = \pa + \cB_n \( \pa - \cB_{n} - \cA_{n} \)^{-1} \cA_n
\lab{wy}
\ee
which, upon the identification $\cB_n  = \bsj\, ,\,\cA_n = \sj $, takes the
form $L= D + \bsj\, \(D - \sj \,- \bsj \,\)^{-1} \sj$. This is the form
of the two-boson KP hierarchy which appeared in connection
with the $SL(2,\IR)/U(1)$ coset model \ct{YW9111}.

The above simple 2-boson model will now be generalized
to the arbitrary multi-boson KP hierarchies.

We start by finding a ``square-root'' lattice formulation corresponding to
the general Toda lattice hierarchy \rf{spectr}.
In this spirit we are led to a spectral equation:
\br
\l^{1/2}\; {\wti \Psi}_{n+\h} \eq \Psi_{n+1} + \cA^{(0)}_{n+1} \Psi_n +
\sum_{p=1}^M \cA^{(p)}_{n-p+1} \Psi_{n -p} \lab{sqraa} \\
\l^{1/2} \; \Psi_n \eq {\wti \Psi}_{n+\h} + \cB^{(0)}_n {\wti \Psi}_{n-\h}
\lab{sqrbb}
\er
with time evolution equations:
\be
{\wti \Psi}_{n+\h} = \( \pa - \cB^{(0)}_n - \cA^{(0)}_{n} \){\wti \Psi}_{n-\h}
\quad;\quad
\Psi_{n+1} = \( \pa - \cB^{(0)}_n - \cA^{(0)}_{n+1} \) \Psi_{n}
\lab{evolution}
\ee
As in the two-boson case above, it is straightforward to show that, upon
excluding the half-integer modes, the generalized Volterra system
\rf{sqraa}--\rf{evolution} reduces to the Toda lattice
spectral eqs.\rf{spectr} for $M+1$, where:
\br
a_0 (n) = \cA^{(0)}_{n+1} + \cB^{(0)}_n \quad ; \quad
a_{M+1}(n) = \cB^{(0)}_n \cA^{(M)}_{n-M}  \lab{Tod-Volt} \\
a_p (n) = \cA^{(p)}_{n-p+1} + \cB^{(0)}_n \cA^{(p-1)}_{n-p+1} \quad , \;\;
p=1,\ldots ,M \nonu
\er
{}From \rf{sqraa}--\rf{evolution} we find:
\be
\l^{1/2}\; {\wti \Psi}_{n+\h}= \( \pa - \cB^{(0)}_n + \sum_{p=1}^M
\cA^{(p)}_{n-p+1} ( \pa - \cB^{(0)}_{n-p} - \cA^{(0)}_{n-p+1} )^{-1}
\cdots ( \pa - \cB^{(0)}_{n-1} - \cA^{(0)}_{n} )^{-1} \) \Psi_n
\lab{speca}
\ee
and
\be
\l^{1/2}\; \Psi_{n}= \( \pa - \cA^{(0)}_n \) {\wti \Psi}_{n-\h}
\lab{specb}
\ee
{}From the last two relations it follows that:
\br
\l\; \Psi_{n}\eq\( \pa - \cA^{(0)}_n \)
\( \pa - \cB^{(0)}_{n-1} + \sum_{p=1}^M
\cA^{(p)}_{n-p} ( \pa - \cB^{(0)}_{n-p-1} - \cA^{(0)}_{n-p} )^{-1}
\cdots ( \pa - \cB^{(0)}_{n-2} - \cA^{(0)}_{n-1} )^{-1} \) \nonu \\
&&( \pa - \cB^{(0)}_{n-1}-\cA^{(0)}_{n} )^{-1} \Psi_n
\lab{almost}
\er
This defines a Lax operator through $ \l \Psi_n =L^{(M+1)}_n
\Psi_{n}$ where
\be
L^{(M+1)}_n = e^{\int \cB^{(0)}_{n-1}} \( \pa - \cA^{(0)}_n +\cB^{(0)}_{n-1}\)
L^{(M)}_n \( \pa - \cA^{(0)}_n \)^{-1} e^{-\int \cB^{(0)}_{n-1}}
\lab{recu}
\ee
and
\be
L^{(M)}_n=  \pa  + \sum_{p=1}^M
\cA^{(p)}_{n-p} ( \pa + \cB^{(0)}_{n-1}- \cB^{(0)}_{n-p-1} -
\cA^{(0)}_{n-p} )^{-1}
\cdots ( \pa + \cB^{(0)}_{n-1}-\cB^{(0)}_{n-2} - \cA^{(0)}_{n-1} )^{-1}
\lab{lax}
\ee
Eqs. \rf{recu}--\rf{lax} can be identified with the recurrence
relation for the $2M$-boson KP Lax operators \rf{3-3}
established in \ct{ANP9401} :
\br
L_{M+1} = e^{\int c_{M+1}} \( D + c_{M+1} - e_{M+1} \) L_M
\( D - e_{M+1} \)^{-1} e^{-\int c_{M+1}} \lab{1a} \\
M=0,1,2 \ldots \quad , \quad L_0 \equiv D \phantb \nonu \\
\lcurl c_k(x) \, ,\, e_l(y) \rcurl = - \d_{kl} \pa_x \d (x-y) \quad , \;
k,l =1,2, \ldots, M+1 \lab{1b}
\er
The free-field pairs $(c_r ,e_r )_{r=1}^{M+1}\,$ are the ``Darboux-Poisson''
canonical pairs for
the second KP bracket \rf{second-KP} satisfied by $L_{M+1}$ for arbitrary
$M$ \ct{ANP9401}.
This defines a sequence of the multi-boson KP Lax operators in terms of
Darboux-Poisson free-field pairs with respect to the second KP bracket, very
much like \rf{3-1} defined a similar sequence of Lax operators in terms of
Darboux-Poisson free-field pairs with respect to the first KP bracket
\ct{ANP93}.
This construction can be viewed as a generalized Miura transformation
for multi-boson KP hierarchies, and  hence ``abelianization'' of the
{\em second} KP Hamiltonian structure \rf{second-KP}, {\sl i.e.},
expressing the coefficient fields of the pertinent KP Lax operator
in terms of canonical pairs of free fields.

Eq.\rf{1a} implies the following recurrence relations for the coefficient
fields of $L_M$ \rf{3-3} (see also \ct{ANP9401}) :
\br
B^{(M+1)}_k = B^{(M)}_k + c_{M+1} \quad 1 \leq k \leq M \quad ,
B^{(M+1)}_{M+1} = c_{M+1} + e_{M+1}  \lab{2a}  \\
A^{(M+1)}_1 = \( \pa + B^{(M)}_1 + c_{M+1} - e_{M+1} \) A^{(M)}_1 \lab{3a} \\
A^{(M+1)}_k = A^{(M)}_{k-1} +
\( \pa + B^{(M)}_k + c_{M+1} - e_{M+1} \) A^{(M)}_k
\quad ,\quad 2 \leq k \leq M   \lab{4a} \\
A^{(M+1)}_{M+1} = A^{(M)}_M + \( \pa + c_{M+1}\) e_{M+1} \qquad \quad \lab{5a}
\er

As found in ref.\ct{ANP9401}, the recurrence relations \rf{2a}--\rf{5a} have
an explicit solution:
\br
B^{(M)}_k \eq e_k +\sum_{l = k}^{M}  c_l \quad , \quad
1 \leq k \leq M \qquad A^{(M)}_{M} =\sum_{k=1}^{M}
\( \pa +c_k \) e_k \lab{11}  \\
A^{(M)}_k \eq \sum_{n_{M-k+1} =1}^k \( \pa + e_{n_{M-k+1}} - e_{n_{M-k+1}+M-k}
+ \sum_{l_k = n_{M-k+1}}^{n_{M-k+1}+M-k} c_{l_k} \) \times \nonu  \\
&\times& \!\!  \sum_{n_{M-k}=1}^{n_{M-k+1}} \( \pa + e_{n_{M-k}} - e_{n_{M-k}
+M-1-k}+ \sum_{l_{k-1} = n_{M-k}}^{n_{M-k}+M-1-k} c_{l_{k-1}} \)
\times \cdots  \lab{14}\\
&\times& \!\! \sum_{n_2 =1}^{n_3} \( \pa + e_{n_2} - e_{n_2 +1} +
c_{n_2} + c_{n_2 +1} \) \, \sum_{n_1 =1}^{n_2} \( \pa + c_{n_1} \) e_{n_1}
\quad , \;\; k=1, \ldots, M-1 \nonu
\er
in terms of the free fields $(c_r ,e_r )_{r=1}^{M}\,$.

A simple calculation, based on the explicit expressions \rf{11}-\rf{14},
gives the following general relations (valid for any $M$):

\prop
\br
\pbr{B^{(M)}_i (x)}{B^{(M)}_j (y)} \eq - X_{ij}\, \pa_x \d (x-y)
\qquad; \qquad X_{ij} \equiv \d_{ij} + 1 \lab{xdef} \\
\pbr{A^{(M)}_{M}(x)}{B^{(M)}_k (y)}\eq  - \((M+1-k) \pa_x + B^{(M)}_k\)
\pa_x \d (x-y) \;\; , \;\; 1 \leq k \leq M \nonu \\ 
\pbr{A^{(M)}_{M}(x)}{A^{(M)}_{M}(y)} \eq - \( A^{(M)}_{M} (x) \pa_x +
\pa_x {A^{(M)}_{M}}\)  \d (x-y)\nonu \\ 
\pbr{A^{(M)}_{i}(x)}{B^{(M)}_j (y)}\eq  0    \qquad {\rm for}\; i < j\nonu
\er
where for brevity we only recorded the most simple relations
following from \rf{11}-\rf{14}.
\lskip
(1) \underbar{Example} -- 2-boson KP :
\br
L_1 &=& e^{\int c_1} \( D + c_1 - e_1 \) D \( D - e_1 \)^{-1} e^{-\int c_1} =
D + A^{(1)}_1 \( D - B^{(1)}_1 \)^{-1} \lab{7} \\
&&A^{(1)}_1 = \( \pa + c_1\) e_1 \quad ,\quad  B^{(1)}_1 = c_1 + e_1
\lab{8}
\er
Here we recognize the structure of the two-boson hierarchy from \rf{celax} as
well the generalized Miura map \rf{gmiura}.
\lskip
(2) \underbar{Example} -- 4-boson KP :
\br
L_2 &= &e^{\int c_2} \( D + c_2 - e_2 \)
\lb D + A^{(1)}_1 \( D - B^{(1)}_1 \)^{-1} \rb
\( D - e_2 \)^{-1} e^{-\int c_2}  \nonu  \\
&= &D + A^{(2)}_2 \( D - B^{(2)}_2 \)^{-1}
+ A^{(2)}_1 \( D - B^{(2)}_1 \)^{-1} \( D - B^{(2)}_2 \)^{-1}  \lab{9} \\
A^{(2)}_2 &= &A^{(1)}_1 + \( \pa + c_2\) e_2  =
\( \pa + c_1\) e_1 + \( \pa + c_2 \) e_2     \lab{10a} \\
A^{(2)}_1 &=& \( \pa + B^{(1)}_1 + c_2 - e_2 \) A^{(1)}_1
= \( \pa + e_1 + c_1 + c_2 -e_2 \) \( \pa + c_1 \) e_1  \lab{11a} \\
B^{(2)}_2 &=& c_2 + e_2  \quad , \quad B^{(2)}_1 = B^{(1)}_1 + c_2
= e_1 + c_1 + c_2   \lab{10b}
\er
where $A^{(1)}_1$ and $B^{(1)}_1$ are substituted with their expressions
\rf{8}. It is easy to derive the second bracket structure for the above fields
directly from \rf{1b}.

$\phantom{aaaa}$\par
{}From the recursive relation \rf{1a} we can obtain
closed expressions for the general Lax operator $L_{M}$, $M=1,2,\ldots$,
directly in terms of the building blocks $(c_k,e_k)_{k=1}^{M}$ :
\br
L_{M} \eq  \( D - e_{M} \) \prod_{k=M-1}^1 \( D - e_k -
\sum_{l=k+1}^{M} c_l \) \( D - \sum_{l=1}^{M} c_l \) \prod_{k=1}^{M}
\( D - e_k - \sum_{l=k}^{M} c_l \)^{-1}  \nonu \\
\eq \prod_{k=M}^1 \( D + c_k - B_k \) \( D - \sum_{l=1}^{M} c_l \)
\prod_{k=1}^{M} \( D - B_k\)^{-1}  \lab{1c}
\er
where for brevity we drop from now on the superscript $M$, so
that $B_k \equiv B^{(M)}_k =e_k + \sum_{l=k}^{M} c_l$.
Let us introduce now the new linear combinations:
\be
\bb_i \equiv c_{M-i+1} - B_{M-i+1} \quad;\quad
\bb_{M+1} \equiv - \sum_{l=1}^M c_{l} \qquad i=1, \ldots , M
\lab{bbar}
\ee
The advantage of this notation is that the Lax operator $L_M $ \rf{1c} takes
a more compact form:
\be
L_M = \prod_{j=1}^{M+1} \( D + \bb_j \) \prod_{k=1}^{M} \( D - B_k\)^{-1}
\lab{1cc}
\ee
This form of the Lax operator has already appeared in \ct{fyu}.
The fields (currents) $\bb_j, B_i$ satisfy by definition the condition
\be
\psi_{M+1,M} \equiv \sum_{j=1}^{M+1} \bb_j + \sum_{k=1}^{M}  B_k =0
\lab{gradtrac}
\ee
recognized in \ct{fyu} as a tracelessness condition of the graded
$SL (M+1,M)$ Kac-Moody algebra.
It follows from \rf{xdef} and from:
\be
\{ c_i (x)\, ,\, B_j  (y)\} = -\d_{ij}  \d^{\pr} (x-y)
\lab{cbbra}
\ee
that the Poisson bracket algebra satisfied by the fields $\bb_j, B_i$
is the Cartan subalgebra of the graded $SL (M+1,M)$ Kac-Moody
algebra:
\br
\{ \bb_i (x)\, ,\, \bb_j (y)\} \eq  \(\d_{ij} -  1\) \, \d^{\pr} (x-y) \quad
\quad i,j =1,\ldots, M+1 \nonu\\
\{ B_k (x) \, ,\, B_l (y) \} \eq -\(\d_{kl} + 1 \) \d^{\pr} (x-y) \quad \quad
k,l=1, \ldots,M  \lab{bbbij}\\
\{ \bb_i (x)\, ,\, B_l (y) \}\eq  \d^{\pr} (x-y)
\nonu
\er
We shall refer to the algebra \rf{bbbij} as $SL_{c} (M+1,M)$
Kac-Moody algebra.

\sect{Reduction of ${\bf SL (M+1,M) }$ to ${\bf SL (M+1,M-k) }$ }
Here we present a general scheme of gauging away $k$ (out of $M$)
currents $B_{M-k+1}, \ldots, B_M$ by introducing the gauge generator
\be
G \, = \, \exp \( - \int \sum_{i=1}^M B_i \g_i \)
\lab{gmboson}
\ee
which induces the following transformations via Hamiltonian action (as in
\rf{gjg1}) :
\br
B_i &\to& {\wti B}_i = G^{-1} B_i G = B_i + X_{ij} \g_j^{\pr}\nonu  \\
c_i &\to& {\wti c}_i = G^{-1} c_i G = c_i + \g_i^{\pr} \lab{bitrans} \\
e_i &\to& {\wti e}_i = G^{-1} e_i G = e_i + \sum_{l=1}^i \g_l^{\pr} \nonu
\er
Note that $ \g_i^{\pr} $ are fixed by the gauge-fixing condition
$ {\wti B}_i = 0 \;, \; i = M-k+1, \ldots, M\,$ i.e.
\be
\g_i^{\pr} = - {X^{(k)}_{ij}}^{-1} B_j \qquad ; \qquad X^{(k)}_{ij}
\equiv 1 + \d_{ij}
\lab{xkdef}
\ee
with $X^{(k)}_{ij}$ being restriction of $X_{ij}$ to
$M-k+1 \leq i,j \leq M$ and $ \g_r^{\pr}=0 $ for $ 1 \leq r \leq M-k$.

In what follows we shall need to find explicitly the inverse of the matrix
$X_{ij} \equiv 1 + \d_{ij} $.
Let ${\bf U}$ be a $M \times M$ matrix with elements $U_{ij} =1$. Hence
${\bf X}= \one + {\bf U} $ and it is easy to see
that:
\be
{\bf X}^{-1} = \one - {\bf U} + {\bf U}^2 - \ldots = \one - {\bf U}
\( 1 -M +M^2 - M^3 +\ldots \) = \one - { 1\over 1 +M} {\bf U}
\lab{xinv}
\ee
Correspondingly we also find
\be
{\bf X^{(k)}}^{-1} =  \one - { 1\over 1 +k } {\bf U^{(k)}}
\lab{xkinv}
\ee
where again the superscript $(k)$ indicates restriction of the matrix indices
to $M-k+1 \leq i,j \leq M$ .

With this information we can rewrite \rf{xkdef} as:
\be
\g_i^{\pr} = -  B_i + {1\over {k+1}} \sum_{n=M-k+1}^M B_n
\quad , \;\; M-k+1 \leq i \leq M  \lab{gamma}
\ee
{}From \rf{gamma} and \rf{bitrans} we find the values of the gauge-rotated
non-zero $B_i$ to be given by:
\be
{\wti B}_r = B_r - {1\over {k+1}} \sum_{n=M-k+1}^M B_n
\quad , \;\; 1 \leq r \leq M-k    \lab{tibr}
\ee
For the gauge transformed $ c_i$ we find:
\be
{\wti c}_i = \left\{ \begin{array}{ll}  c_i & \quad 1 \leq i \leq M-k \\
\phantom{aaaaa} & \phantom{aaaa} \\
c_i - B_i + {1\over {k+1}} \sum_{n=M-k+1}^M B_n  & \quad M-k+1 \leq i \leq M
 \end{array} \right.
 \lab{ciarray}
 \ee
Correspondingly, we obtain for the new gauge transformed
$\bb_i=\( {\wti c}_{M-i+1} - {\wti B}_{M-i+1}, - \sum_{l=1}^M {\wti c}_l \) $
\rf{bbar} (omitting the tilde on top of $\bb_i$ for brevity) :
\be
{\bb}_i = \left\{ \begin{array}{ll}
c_{M-i+1} - B_{M-i+1} + {1\over {k+1}} \sum_{n=M-k+1}^M B_n  &
\quad 1 \leq i \leq M \\
\phantom{aaaaa} & \phantom{aaaa} \\
- \sum_{l=1}^M c_l + {1\over {k+1}} \sum_{n=M-k+1}^M B_n
&\quad i = M+1    \end{array} \right.
 \lab{bbiarray}
 \ee
Using \rf{xdef} and \rf{cbbra} we find that $\bb_i, {\wti B}_r$ satisfy the
Poisson-bracket Cartan subalgebra of the graded $SL (M+1, M-k)$ Kac-Moody
algebra (cf. eqs.\rf{bbbij}) :
\br
\{ \bb_i (x)\, ,\, \bb_j (y) \} \eq  \(\d_{ij} -  {1 \over k+1}\) \d^{\pr}
(x-y)
\quad\quad i,j =1,\ldots M+1 \nonu\\
\{ {\wti B}_r (x)\, ,\, {\wti B}_s (y) \} \eq
-\(\d_{rs} + {1 \over k+1} \) \d^{\pr}(x-y)
\quad\quad 1 \leq r,s \leq M-k  \lab{redbbb}\\
\{ \bb_i (x) \, ,\, {\wti B}_r (y) \}\eq  - {1 \over k+1}\, \d^{\pr} (x-y)
\nonu
\er
for which we shall use the symbol $SL_c (M+1, M-k)$.
Note that the gauge transformation \rf{bitrans} maps the trace-constraint
$\psi_{M+1,M}$ \rf{gradtrac} to the new $SL (M+1, M-k)$ trace condition:
\be
\psi_{M+1,M-k} \equiv  G^{-1}\, \psi_{M+1,M}\, G =\sum_{j=1}^{M+1} \bb_j +
\sum_{r=1}^{M-k} {\wti B}_r = 0
\lab{psimk}
\ee
In addition to $\bb_i, {\wti B}_r$
there are $k$ gauge parameters $\g_n^{\pr}$ \rf{gamma}
associated with the $SL (k+1)$ algebra and satisfying:
\be
\{ \g_n^{\pr}(x) \, ,\, \g_m^{\pr}(y) \}
= - \(\d_{nm} -  {1 \over k+1}\) \d^{\pr} (x-y)
\quad\quad n,m =M-k+1,\ldots, M   \lab{gleft}
\ee
which we shall call $SL_c (k+1)$ algebra.
Note that $\g_n^{\pr}$ are decoupled from $\bb_i, {\wti B}_r$ :
\be
\lcurl \g_n^{\pr} \, ,\, {\bar B}_i \rcurl =
\lcurl \g_n^{\pr} \, ,\, {\wti B}_r \rcurl = 0 \quad , \quad
n=M-k+1,\ldots ,M \; ,\; r = 1,\ldots ,M-k \; ,\; i = 1,\ldots M+1
\lab{decouple}
\ee
We therefore easily arrive at the following:

\prop
{\em The second bracket of the multi-boson KP is reducible under gauge
fixing procedure described above and $\{ \bb_i \, ,\, B_n \} =
 \{ {\wti B}_r \, ,\,B_n \} =  0$ where
$B_n\; , \, M-k+1 \leq n \leq M$ are the modes gauged away.}

This result generalizes the observation, we made earlier in section 3.2,
concerning the decoupling of the KdV mode $u=a- {1\over 4} b^2 - \h b^{\pr} $
from the current $b$.

The above process of reduction can be extended to remove all
currents, {\sl i.e.}, $ {\wti B}_i = 0 $ for $i=1,\ldots M$. In this case:
\be
\g_i^{\pr} = - X_{ij}^{-1} B_j \qquad ; \qquad
{\wti c}_i = c_i - X_{ij}^{-1} B_j
\lab{ectilde}
\ee
Accordingly $\bb_i= \( {\wti c}_{M-i+1},  - \sum_{l=1}^M c_{l} \)$ and the
underlying Poisson bracket algebra splits into two disjoint $SL_c (M+1)$
algebras of opposite signatures:
\br
\{\bb_i (x)\, ,\, \bb_j (y)\} \eq  \(\d_{ij} - { 1\over 1 +M}\)
\d^{\pr} (x-y)\quad i,j=1,\ldots M \nonu\\
\{ \g_i^{\pr}(x) \, ,\, \g_j^{\pr}(y) \} \eq - \(\d_{ij} - { 1\over 1 +M}\)
\d^{\pr} (x-y)  \quad  i,j=1,\ldots M \lab{ticalpha}\\
\{ \bb_i (x)\, ,\, \g_j^{\pr}(y) \}\eq  0
\nonu
\er

\underbar{Example} -- 2-boson KP :
Recall expressions \rf{gmiura}, $A= e^{\pr}+ ce\, , \, B= c + e$.
In this case: ${\wti B}  = B + 2\g^{\pr}\, , \,{\wti c}  = c + \g^{\pr}$
and condition ${\wti B}  = 0$ leads to ${\wti c}  = (c -e)/2$ and
$\g^{\pr}= - (c +e)/2$.
Correspondingly we find:
$A \to {\wti A}= {\wti e}^{\pr}+ {\wti c}{\wti e}= (e -c)^{\pr}/2 -
(e -c)^{2}/4$, which satisfies the Virasoro algebra due to
$ \pbr{(e -c)/2}{(e -c)/2}= - \d^{\pr} (x-y)/2$.
Note also that we can rewrite $ {\wti A}$ as $ A- {1\over 4} B^2 - \h B^{\pr}$
obtaining agreement with the result of the DS reduction \rf{udef}.

$\phantom{aaa}$ \par
The above considerations show that the graded $SL_c (M+1,M)$ algebra is
reducible and the splitting is of the form:

\prop {\em $SL_c (M+1,M)= SL_c (M+1,M-k) \oplus SL_c (k+1)$ with
$k=1,\ldots, M$}

Indeed, given $\{ {\bar \b}_i, \b_r\}$ with $1 \leq i \leq M+1\, , \,
1  \leq r \leq M-k $ spanning $SL_c (M+1,M-k)$ and an independent
basis $\{ \a_n\}$ for $SL_c (k+1)$ with $M-k+1 \leq n \leq M $,
we can form an $SL_c (M+1,M)$  algebra by taking the following linear
combination:
\be
\bb_i = \left\{ \begin{array}{l}  {\bar \b}_n - \a_n + X^{(k)}_{nm} \a_m
\\ \b_r+ X_{rm} \a_m \\
{\bar \b}_{M+1} +\sum_{n=1}^{k}  \a_n     \end{array} \right.
\qquad\; B_i = \left\{ \begin{array}{l}  - X^{(k)}_{nm} \a_m\\
\b_r - X_{rm} \a_m \end{array} \right.
\lab{oplus}
\ee
Note, that this construction, as we have seen above, is reversible.
Note also, that for $k=M$ we have a decomposition into
two independent $SL_c (M+1)$ algebras of opposite
signatures as in \rf{ticalpha}.

\subsection{The Second Bracket Structure of the ${\bf SL(M+1,M-k)}$-KdV
Hierarchy}
Let us recall the expression \rf{redbbb} for the graded Poisson bracket
algebra $SL_c (M+1,M-k)$. In \ct{fyu} this algebra was realized as a Dirac
bracket algebra obtained from the Poisson brackets of two independent set of
bose fields of opposite signatures:
\br
\{ \bb_i (x)\, ,\, \bb_j (y)\}_{PB} \eq  \d_{ij} \; \d^{\pr} (x-y)
\quad  \; \;i,j =1,\ldots M+1 \nonu\\
\{ B_m (x) \, ,\, B_n (y) \}_{PB} \eq -\d_{mn} \; \d^{\pr} (x-y) \quad  \; \;
\;
1 \leq m,n \leq M-k  \lab{pbbb}\\
\{ \bb_i (x)\, ,\, B_m (y)\}_{PB} \eq  0
\nonu
\er
by imposing the constraint:
\be
\psi_{M+1,M-k} = \sum_{j=1}^{M+1} \bb_j + \sum_{n=1}^{M-k}  B_n = 0
\lab{psimk1}
\ee
which is  second-class due to:
\be
\{ \psi_{M+1,M-k} (x)\, , \, \psi_{M+1,M-k}(y)\} =
(k +1)  \;\d^{\pr} (x-y)
\lab{psimkbra}
\ee

We are interested in describing the corresponding bracket structures
associated with the generalization of the Lax operator \rf{1cc} to:
\be
\cL_{M+1,M-k} \equiv  \prod_{j=1}^{M+1} \( D + \bb_j \)
\prod_{n=1}^{M-k} \( D - B_n\)^{-1}
\lab{1cce}
\ee
In \ct{fyu} it was pointed out that the Lax operator $\cL_{M+1,M-k} $ (with no
trace condition imposed) satisfies
the second Gelfand-Dickey bracket provided the fields $\bb_i, B_n$ satisfy
\rf{pbbb}. The proposition is therefore:

\prop
\br
&&{\pbbr{\me{\cL_{M+1,M-k}}{X}}{\me{\cL_{M+1,M-k}}{Y}}}_{PB} = \nonu\\
 &&{\Tr}_A \( \( \cL_{M+1,M-k}X\)_{+} \cL_{M+1,M-k}Y -
\( X\cL_{M+1,M-k}\)_{+} Y\cL_{M+1,M-k} \)
\lab{lprop}
\er
where the subscript $PB$ stands for the Poisson bracket as defined by
\rf{pbbb}.
The statement can easily be proved by induction in $M$ and $k$.
It is straightforward
to verify \rf{lprop} for $M=0,k=0$ just by inserting $\cL_{1,0}
= \( D + \bb \)$ into the formula \rf{lprop}.
The essential part of the induction proof with respect to $M$
consists in showing
that \rf{lprop} is valid for $\cL_{M+1} = ( D + \bb) \cL_{M}$
provided it is valid for $\cL_{M}$ with $k=0$. We have:
\br
&&\pbbr{\me{\cL_{M+1}}{X}}{\me{\cL_{M+1}}{Y}}_{PB} \Bgv_{{\bar B}={\rm fixed}}
= \pbbr{\me{\cL_{M}}{X ( D + \bb)}}{\me{\cL_{M}}{Y( D + \bb)}}_{PB}
\Bgv_{{\cL}_M = {\rm fixed}} \nonu\\
&+& \pbbr{\me{( D + \bb)}{\cL_M X}}{\me{( D + \bb)}{\cL_M Y}}_{PB} \nonu \\
\eq
{\Tr}_A \( \( \cL_{M}X( D + \bb)\)_{+} \cL_{M}Y( D + \bb) -
\( X( D + \bb)\cL_{M}\)_{+} Y( D + \bb)\cL_{M} \) \nonu\\
&+&\int dx\, {\rm Res}(\cL_{M}X) \,  \pa_x {\rm Res}(\cL_{M}Y)
\lab{pb-B}
\er
Now, using the simple identity for pseudo-differential operators
(valid for any $\bb$) :
\be
( D + \bb) \( \cL_{M}X( D + \bb)\)_{+} =
 \( ( D + \bb)\cL_{M}X\)_{+} ( D + \bb)
+  \pa \, {\rm Res}(\cL_{M}X)
\lab{plusplus}
\ee
we arrive at the desired result:
\be
{\pbbr{\me{\cL_{M+1}}{X}}{\me{\cL_{M+1}}{Y}}}_{PB} =
{\Tr}_A \( \( \cL_{M+1}X\)_{+} \cL_{M+1}Y-
\( X\cL_{M+1}\)_{+} Y\cL_{M+1} \)
\ee
The remaining step of the induction proof with respect to $k$,
involving the transition $\cL_{M+1,k} \to \cL_{M+1,k+1}$,
can be performed using the same techniques as above.

We now turn our attention to the Dirac bracket which results from \rf{lprop}
by imposing the constraint \rf{psimk1}. The relevant statement is:

\prop
\br
&&{\pbbr{\me{\cL_{M+1,M-k}}{X}}{\me{\cL_{M+1,M-k}}{Y}}}_{DB} =
{\pbbr{\me{\cL_{M+1,M-k}}{X}}{\me{\cL_{M+1,M-k}}{Y}}}_{PB} \nonu\\
&+&{1 \over k+1} \int dx \; {\rm Res}\( \sbr{\cL_{M+1,M-k}}{X}\) \pa^{-1}
{\rm Res}\( \sbr{\cL_{M+1,M-k}}{Y}\)
\lab{ldb}
\er
The proof follows from the computation of the extra term of the relevant Dirac
bracket:
\br
&-&\int {\pbbr{\me{\cL_{M+1,M-k}}{X}}{\psi_{M+1,M-k}}}_{PB}
{\pbbr{\psi_{M+1,M-k}}{\psi_{M+1,M-k}}}_{DB}^{-1}\times \nonu \\
&\times&{\pbbr{\psi_{M+1,M-k}}{\me{\cL_{M+1,M-k}}{Y}}}_{DB} \lab{extra}
\er
One easily verifies the equality of \rf{extra} with the second term on the
r.h.s. of \rf{ldb} using \rf{psimkbra}:
\br
 {\pbbr{\me{\cL_{M+1,M-k}}{X}}{\psi_{M+1,M-k}(z)}}_{PB} \eq
- \me{\sbr{\d (x-z)}{\cL_{M+1,M-k}}}{X} \nonu \\
\eq  -{\rm Res}\( \sbr{X}{\cL_{M+1,M-k}}\)(z)
\er

Formula \rf{ldb} contains as special cases the KP hierarchy, corresponding to
$k=0$, and the KdV hierarchy, corresponding to $k=M $ (see e.g. \ct{diz}).
For the intermediary cases $ 0 < k < M$ eq.\rf{ldb} represents a compact
expression for the Poisson bracket structure of all
$SL (M+1,M-k)$-KdV hierarchies defined by the Lax operators \rf{1cce}.

\subsection{The Lax Formulation of ${\bf SL (M+1, M-k)}$-KdV}
In this subsection we give expressions for the coefficients of the Lax
operators of the generalized $ SL (M+1, M-k)$-KdV hierarchy in terms of free
fields.
Let us start with the case where all the currents $B_i$ are gauged away
according to \rf{ectilde}.
In this limit the expression \rf{14} becomes:
\br
{\wti A}_M \equiv {\wti A}^{(M)}_{M} \eq \sum_{k=1}^{M}
\( \pa +{\wti c}_k \) \( - \sum_{l =k}^M {\wti c}_{l}\)\lab{11b}  \\
{\wti A}_k \equiv {\wti A}^{(M)}_k \eq \sum_{n_{M-k+1} =1}^k \( \pa +
{\wti c}_{n_{M-k+1}+M-k} \) \times
  \sum_{n_{M-k}=1}^{n_{M-k+1}} \( \pa + {\wti c}_{n_{M-k}+M-1-k}\)
\times \cdots  \nonu  \\
&\times& \!\! \sum_{n_2 =1}^{n_3} \( \pa +
{\wti c}_{n_2 +1} \) \, \sum_{n_1 =1}^{n_2} \( \pa + {\wti c}_{n_1} \)
\( - \sum_{l =n_1}^M {\wti c}_{l}\)
\quad , \;\; k=1, \ldots, M-1 \lab{ti14}
\er
which agrees with Fateev-Lukyanov \ct{FL88} expression:
\br
&& \prod_{i=1}^{M+1} \( D  + \bb_{i} \)
= D^{M+1} + {\wti A}_M D^{M-1} + \ldots + {\wti A}_1
\lab{fateev} \\
&& \bb_{M+1} = - \sum_{l =1}^M {\wti c}_{l} \quad;\quad \bb_i =
{\wti c}_{M+1 -i} \;\;  ; \;\; i = 1,\ldots,M \nonu
\er
The reason for this agreement is that the Lax operator in \rf{fateev} is equal
to $\cL_{M+1,0}$ (eq.\rf{1cce} for $k=0$)
with the condition $\psi_{M+1,0}=0$ imposed.
Both approaches (Dirac bracket and gauge fixing) lead to the algebra
\rf{ticalpha} or, equivalently, to the formula \rf{ldb} with $k=M$.
Hence, our construction has provided a simple proof for the
Fateev-Lukyanov expression \ct{FL88} (see also \ct{Dickey,KW81}).
Note that the use of the gauge fixing method has the advantage that
the Lax coefficients \rf{ti14} of the $SL(M+1)$-KdV
can still be written in terms of the original Darboux-Poisson pairs
$(c_r ,e_r )_{r=1}^{M}$ through ${\wti c}_i$ from \rf{ectilde}
which is given by :
\be
{\wti c}_i = - e_i - \sum_{l=i+1}^M c_l + {1 \over M+1} \sum_{l=1}^M \( e_l + l
c_l \) \qquad i =1, \ldots, M
\lab{wtic}
\ee
Generally, to obtain a convenient expression for the Lax operator
of the $SL(M+1, M-k)$-KdV
associated with the graded $SL (M+1, M-k)$ algebra,
we gauge away $B_M, \ldots ,B_{M-k+1}$ leaving ${\wti B}_{1}, \ldots,
{\wti B}_{M-k}$.
We obtain in this way from \rf{1c} :
\be
\cL_{M+1,M-k} = \prod_{l=M}^{M-k+1} \( D +  {\wti c}_{l}\)
\prod_{l=M-k}^{1} \( D +  {\wti c}_{l}- {\wti B}_{l}\)
\( D - \sum_{l=1}^M {\wti c}_{l} \)\prod_{l=1}^{M-k}
\( D - {\wti B}_{l}\)^{-1}
\lab{laxmk}
\ee
which automatically satisfies the appropriate trace condition \rf{psimk1}.
We can interpret \rf{laxmk} as a superdeterminant of the graded
$SL (M+1, M-k)$ matrix in a diagonal gauge, which for the ordinary
KdV case $k=M$ becomes an ordinary determinant as in  \rf{fateev}.

We can alternatively rewrite the last expression \rf{laxmk} in a way,
which corresponds to the DS gauge as:
\be
\cL_{M+1,M-k}
= \sum_{l=1}^{M-k} {\wti A}_l  \prod_{i=l}^{M-k} \( D - {\wti B}_{i}\)^{-1}
+ \sum_{l=0}^{k-1} {\wti A}_{l+M-k+1}  D^{l} + D^{k+1} \lab{1ff}
\ee
with the second bracket structure automatically given by formula \rf{ldb}.
The coefficients ${\wti A}_l$ can be explicitly expressed in terms of
${\wti c}_l , {\wti B}_l$ from \rf{11}-\rf{14} by substituting there
\be
c_l \to {\wti c}_l \quad ;\quad
e_l \to {\wti B}_l - \sum_{i=l}^M {\wti c}_i \;\; ,\; l=1,\ldots ,M-k
\;\; ,\;\; e_l \to - \sum_{i=l}^M {\wti c}_i \;\; ,\; l=M-k+1,\ldots ,M
\lab{subst}
\ee
Hence, again we arrived at representation of the coefficients of the
$SL(M+1,M-k)$-KdV Lax operators \rf{1ff}, or \rf{laxmk},
in terms of the free fields (currents)  whose Poisson bracket algebra is
given by \rf{redbbb}
(recall ${\bar B}_i = {\wti c}_{M-i+1}$ for
$1 \leq i \leq M-k$ and
${\bar B}_i = {\wti c}_{M-i+1} - {\wti B}_{M-i+1}$ for $M-k+1 \leq i \leq M$).
Correspondingly, the Lax coefficients ${\wti A}_i , {\wti B}_j$ in
\rf{1ff} satisfy a nonlinear Poisson bracket algebra, called
$W(M,M-k)$-algebra in ref.\ct{BX9311}, which results from \rf{ldb}.
This $W(M,M-k)$-algebra is a generalization of the well-known Zamolodchikov's
nonlinear $W_{M+1}$ algebra \ct{Zam}, in particular, $W(M,0) \simeq W_{M+1}$.
Thus, eqs.\rf{11}--\rf{14} with the
substitutions \rf{subst} provide explicit free-field realization of
$W(M,M-k)$ .

The Lax operator $\cL_{M+1,M-k}$ \rf{laxmk} (or \rf{1ff}) possesses the
following pseudo-differential series expansion:
\br
\cL_{M+1,M-k} = D^{k+1} +
\sum_{l=0}^{k-1} {\wti A}_{l+M-k+1}({\wti c},{\wti B})  D^{l}
+ \sum_{n=0}^{\infty} w_n ({\wti c},{\wti B}) D^{-1-n}  \lab{series} \\
w_n ({\wti c},{\wti B}) = \sum_{r=0}^{\min (M-k-1,n)} (-1)^{n-r}
{\wti A}_{M-k-r} ({\wti c},{\wti B}) P^{(r+1)}_{n-r} \( -{\wti B}_{M-k-r},
\ldots ,-{\wti B}_{M-k}\)   \lab{w-n}
\er
Here, as above, ${\wti A}_{s}({\wti c},{\wti B})$ are given by the expressions
\rf{11}--\rf{14} with the substitutions \rf{subst}, whereas
\be
P^{(r)}_n \( \phi_1 ,\ldots ,\phi_r \) \equiv \sum_{m_1 + \cdots + m_r =n}
\( \pa + \phi_1 \)^{m_1} \ldots \( \pa + \phi_r \)^{m_r}\cdot 1  \lab{faa}
\ee
denote the multiple \faa polynomials (cf. \ct{ANP93,ANP9401}
for analogous
to \rf{series}-\rf{faa} expressions for the multi-boson KP Lax operators).
The Poisson bracket algebra of the coefficient fields
${\wti A}_{s}({\wti c},{\wti B})\; ,\; s=M-k+1,\ldots ,M$ , and
$w_n ({\wti c},{\wti B}) \; ,\; n=0,1,2,\ldots $, which results from the
free-field Poisson brackets of their constituents \rf{redbbb},
is a nonlinear algebra ${\bf {\hat W}_{\infty}^{(k)}}$ generalizing the
known nonlinear $\hWinf$ algebra \ct{Wu-Yu} (see also \ct{fyu}). In
particular, $\bf {\hat W}_{\infty}^{(k=0)} \simeq \hWinf$.

$\phantom{aaa}$ \par
\underbar{Example} -- $SL(3,1)$-KdV hierarchy.

It is defined by the Lax operator $\, \cL_{3,1} =
{\wti A}_1 \( \pa - {\wti B}_1 \)^{-1} + {\wti A}_2 + \pa^2\,$ (cf. \rf{1ff})
where:
\br
{\wti A}_1 = \( \pa + {\wti B}_1 + {\wti c}_2 \) \(\pa + {\wti c}_1\)
\( {\wti B}_1 - {\wti c}_1 - {\wti c}_2 \)   \lab{A-1}  \\
{\wti A}_2 = \( \pa + {\wti c}_1\) \( {\wti B}_1 - {\wti c}_1\) +
\( {\wti c}_1 - {\wti c}_2 \) {\wti c}_2   \lab{A-2}
\er
and with fundamental Poisson brackets:
\br
\lcurl {\wti c}_1 (x) \, ,\, {\wti B}_1 (y) \rcurl = - \d^{\pr}(x-y)
\quad , \quad
\lcurl {\wti c}_2 (x) \, ,\, {\wti c}_2 (y) \rcurl = \h \d^{\pr}(x-y)
\nonu \\
\lcurl {\wti c}_2 (x) \, ,\, {\wti B}_1 (y) \rcurl = -\h \d^{\pr}(x-y)
\quad ,\quad
\lcurl {\wti B}_1 (x) \, ,\, {\wti B}_1 (y) \rcurl = - {3\over 2} \d^{\pr}(x-y)
\lab{pb-31}
\er
The brackets \rf{pb-31} imply that ${\wti A}_{1,2}$ given by
\rf{A-1},\rf{A-2}, together with ${\wti B}_1$ satisfy the $W(2,1)$ Poisson
bracket algebra \rf{a2alg} (with $\cB \equiv {\wti B}_1$).

$\phantom{aaa}$ \par
Now, one notices the presence of the zero-order term ${\wti A}_{M- k+1}$ in
the Lax operator $\cL_{M+1,M-k}$ \rf{1ff}.
This fact enables us to prove that the $SL (M+1, M-k)$-KdV hierarchy is a
bi-Hamiltonian hierarchy. Consider namely $\cL_{M+1,M-k}^{\pr}=
\cL_{M+1,M-k}- \l $ obtained by redefining the zero-order term in the Lax
operator by addition of the constant $\l$.
Clearly, the second bracket \rf{ldb} for the new Lax operator becomes:
\br
&&\!\!\!{\pbbr{\me{\cL_{M+1,k}^{\pr}}{X}}{\me{\cL_{M+1,k}^{\pr}}{Y}}}_{DB}
\! = {\Tr}_A \( \(\cL_{M+1,k}^{\pr}X\)_{+} \cL_{M+1,k}^{\pr}Y-
\( X\cL_{M+1,k}^{\pr}\)_{+} Y\cL_{M+1,k}^{\pr}\)\nonu\\
&+&{1 \over k+1} \int dx  {\rm Res}\( \sbr{\cL_{M+1,k}^{\pr}}{X}\) \pa^{-1}
{\rm Res}\( \sbr{\cL_{M+1,k}^{\pr}}{Y}\)
- \l \llangle \cL_{M+1,k}^{\pr}\bv \left\lb X,\, Y \right\rb_{R}\rrangle
\lab{pencil}
\er
where we introduced the $R$-commutator $\lb X,\, Y \rb_{R} \equiv
\lb X_{+},\, Y_{+} \rb-\lb X_{-},\, Y_{-}\rb$.
Here again the subscripts $\pm$ denote projections on the pure differential and
pseudo-differential parts of the pseudo-differential operators $X,Y$,
respectively.
Define next an $R$-bracket $ \{\cdot ,\cdot \}_{1}^{R}$
as a bracket obtained from \rf{first-KP} by substituting
$R$-commutator $\lb X,\, Y \rb_{R}$ for the ordinary commutator
\ct{STS83,R82}:
\be
{\pbbr{\me{L}{X}}{\me{L}{Y}}}_1^{R} \equiv
- \llangle L \bv \left\lb X,\, Y \right\rb_{R} \rrangle  \lab{first-RKP}
\ee
Relation \rf{pencil} thus shows that the linear combination of brackets
$\{\cdot ,\cdot \}_{DB} + \l \{\cdot ,\cdot \}_{1}^{R}$ satisfies the Jacobi
identity. We can state this result as:

\prop {\em
${SL (M+1, M-k)}$-KdV hierarchy is bi-Hamiltonian with brackets
$\{\cdot ,\cdot \}_{DB}$ and $\{\cdot ,\cdot \}_{1}^{R}$ defining a compatible
pair of Hamiltonian structures.}

This Proposition establishes, therefore, the fundamental criterium for
integrability of the generalized ${SL (M+1, M-k)}$-KdV hierarchy.

\subsection{The Discrete Symmetry of ${\bf SL (M+1, M-k)}$-KdV Hierarchy.}
Recently, the multi-boson KP hierarchies have  been shown to possess
canonical discrete symmetry realized as a similarity transformation of
their Lax operator \ct{similar}.
It is natural to ask whether the discrete-similarity transformation can be
constructed for the reduced  $SL(M+1, M-k)$-KdV hierarchy for $ k\ne 0$.
One suspects that the presence of $B$ currents in this reduction will allow for
remnants of the discrete symmetry to survive in this system.
We shall show now that this is indeed the case.

First, let us consider
the simplest nontrivial example -- the pseudo-differential Lax operator
of the $SL(3,1)$-KdV hierarchy (for convenience here we
suppress the tildes on the coefficient fields) :
\be
\cL_{3,1} = A_1 { 1 \over {\pa - B_1}} + A_2  + \pa^2
\lab{laxino}
\ee
It is easy to prove its covariance under the similarity transformation:
\be
\( \pa - \cB^0 \)\, \cL_{3,1}\, \( \pa - \cB^0 \)^{-1}
= {\bar A}_1 { 1 \over \pa - \cB^0} + {\bar A}_2  + \pa^2   \lab{simi}
\ee
provided $\cB^0 = B_1 + \pa \ln A_1$ . Eq.\rf{simi} induces
the following discrete transformations on the Lax coefficients which can
be viewed as auto-B\"{a}cklund transformations for the underlying
$SL(3,1)$-KdV hierarchy:
\br
B_1 &\to& {\bar B}_1 = \cB^0 = \cB+ \pa \ln A_1 \lab{discr}\\
A_2 &\to& {\bar A}_2 = A_2 + 2 \pa \(\cB+ \pa \ln A_1\) \nonu \\
A_1 &\to& {\bar A}_1 = A_1 + A_2^{\pr} + \pa \lb
\(\cB+ \pa \ln A_1\)^2 + \pa \(\cB+ \pa \ln A_1\)\rb \nonu
\er
This can be represented by the following Toda-like lattice equations of
motion:
\br
\pa a_2 (n) \eq  a_2 (n)  \lb a_0 ( n+1) - a_0 (n) \rb \lab{lattice}\\
\pa a_0 (n+1) \eq  \h \lb a_1 ( n+1) - a_1 (n) \rb \nonu \\
\pa a_1 (n) \eq  a_2 (n+1) - a_2 (n) - \pa \lb a_0^2 ( n+1) +
\pa a_0 (n+1) \rb \nonu
\er
upon identifying:
\br
a_2 (n) \simeq A_1 \quad ,\quad a_1 (n) \simeq A_2 \quad ,\quad
a_0 (n) \simeq B_1   \lab{simeq} \\
a_2 (n+1) \simeq {\bar A}_1 \quad ,\quad a_1 (n+1) \simeq {\bar A}_2 \quad ,
\quad a_0 (n+1) \simeq {\bar B}_1   \nonu
\er
Eqs.\rf{lattice} can be obtained as consistency conditions of the
following lattice spectral system:
\br
a_2 (n) \Psi_{n-1} + a_1 (n) \Psi_n + \pa^2 \Psi_n \eq \l \Psi_n \nonu \\
\pa \Psi_{n-1} - a_0 (n) \Psi_{n-1} \eq \Psi_{n} \lab{newtoda}
\er

The above discrete symmetry extends to the general case given by the
Lax operator \rf{1ff}.
We find that the similarity transformation
$ \( \pa - \cB^0 \) \cL_{M+1,M-k} \( \pa - \cB^0 \)^{-1}$
with $\cB^0 = B_1 + \pa \ln A_1$ again preserves the form
of the Lax operator, while its coefficients undergo the following
transformations:
\br
B_l &\to& {\bar B}_l = B_{l+1} \lab{discr-a}\\
A_l &\to& {\bar A}_l = A_l + \( \pa + B_{l+1} - \cB^0  \) A_{l+1} \nonu
\er
for the first $M-k-1$ coefficients labelled by  $l =1 , \ldots ,
M-k-1 $ and behaving under similarity transformation in a way consistent
with the lattice-site translations in the underlying Toda-like lattice
\ct{similar} (cf. eqs.\rf{lattice}). For the remaining coefficients we find:
\br
B_{M-k} &\to& {\bar B}_{M-k}= \cB^0 = B_1 + \pa \ln A_1  \lab{discr-b}\\
A_{M-k} &\to& {\bar A}_{M-k} = A_{M-k} + P_{k+1}^{\pr} \(\cB^0 \)
+ \sum_{l=0}^{k-1} \( A_{M-k+l+1} P_{l}\(\cB^0 \) \)^{\pr} \nonu\\
A_{M-k+\a} &\to& {\bar A}_{M-k+\a} =
\sum_{m=1}^{k} \sum_{p=0}^{m-2} { m-1 \choose p}
{m-2-p \choose \a -1} \( A_{M-k+m} P_{p}\(\cB^0 \) \)^{\pr}
P_{m-1-p-\a } \(- \cB^0 \) \nonu \\
&+& \sum_{m=1}^{k} \sum_{p=0}^{m-1} { m-1 \choose p}
{m-1-p \choose \a -1} A_{M-k+m} P_{p}\(\cB^0 \)
P_{m-p-\a } \(- \cB^0 \) \nonu \\
&+& \sum_{p=1}^{k} { k+1 \choose p} {k-p \choose \a -1}
P_{p}^{\pr} \(\cB^0 \) P_{k+1-p-\a } \(- \cB^0 \) \nonu \\
&+& \sum_{p=0}^{k+1} { k+1 \choose p} {k-p+1 \choose \a -1}
P_{p} \(\cB^0 \) P_{k+2-p-\a } \(- \cB^0 \) \nonu
\er
for $1 \leq \a \leq k$. The symbols
$ P_n (\pm \phi) \equiv (D \pm \phi )^n \cdot 1$
denote the ordinary \faa polynomials (cf. \rf{faa}).
Eqs.\rf{discr-a}--\rf{discr-b} represent the auto-B\"{a}cklund
transformations for the generalized $SL(M+1,M-k)$-KdV hierarchies.
Both, the Hamiltonians and Poisson structures of the underlying hierarchies
are invariant under \rf{discr-a}--\rf{discr-b}
due to the similarity character of these auto-B\"{a}cklund transformations.

\sect{Two-Matrix Model as a ${\bf SL (M+1,1)}$-KdV Hierarchy}

Now , let us show how the formalism of the previous sections finds application
in the two-matrix string model.
This model is defined through the partition function:
\be
Z_N \lb t,{\ti t},g \rb = \int dM_1 dM_2 \exp -\lcurl
\sum_{r=1}^{p_1} t_r \Tr M_1^r +
\sum_{s=1}^{p_2} {\ti t}_s \Tr M_2^s + g \Tr M_1 M_2 \rcurl   \lab{2-1}
\ee
Here $M_{1,2}$ are Hermitian $N \times N$ matrices,
and the orders of the matrix ``potentials'' $p_{1,2}$ are assumed to be
finite. In refs.\ct{BX9212,BX9311a} it was shown that, by using the method
of generalized orthogonal polynomials \ct{ortho-poly}, the partition function
\rf{2-1} and its derivatives w.r.t. the parameters $\( t_r,{\ti t}_s ,g \)$
can be explicitly expressed in terms of solutions to constrained generalized
Toda lattice hierarchies associated with \rf{2-1}. The corresponding linear
problem and Lax (or ``zero-curvature'') representation of the latter read
\ct{BX9212,BX9311a} :
\br
Q_{nm} \psi_m = \l \psi_n  \quad , \quad
-g{\bar Q}_{nm} \psi_m = \partder{}{\l} \psi_n   \lab{L-1} \\
\partder{}{t_r} \psi_n = \( Q^r_{(+)}\)_{nm} \psi_m    \quad , \quad
\partder{}{{\ti t}_s} \psi_n = - \( {\bar Q}^s_{-}\)_{nm} \psi_m   \lab{L-2} \\
\partder{}{{\ti t}_s} {\bar Q} = \llb {\bar Q} , {\bar Q}^s_{-}\rrb
\quad , \quad
\partder{}{{\ti t}_s} Q = \llb Q , {\bar Q}^s_{-} \rrb  \lab{L-3} \\
\partder{}{t_r} Q = \llb Q^r_{(+)} , Q \rrb  \quad , \quad
\partder{}{t_r} {\bar Q} = \llb Q^r_{(+)} , {\bar Q} \rrb  \lab{L-4} \\
g \llb Q , {\bar Q} \rrb = \one   \phanta   \lab{string-eq}
\er
The subscripts $-/+$ denote lower/upper
triangular parts, whereas $(+)/(-)$ denote upper/lower triangular plus
diagonal parts.
The parametrization of the matrices $Q$ and ${\bar Q}$ is as follows:
\br
Q_{nn} = a_0 (n) \quad , \quad Q_{n,n+1} =1 \quad ,\quad
Q_{n,n-k} = a_k (n) \quad k=1,\ldots , p_2 -1   \nonu  \\
Q_{nm} = 0 \quad {\rm for} \;\;\; m-n \geq 2 \;\; ,\;\; n-m \geq p_2 \phanta
\lab{param-1}  \\
{\bar Q}_{nn} = b_0 (n) \quad , \quad {\bar Q}_{n,n-1} = R_n \quad , \quad
{\bar Q}_{n,n+k} = b_k (n) R_{n+1}^{-1} \ldots R_{n+k}^{-1}
\quad k=1,\ldots ,p_1 -1    \nonu  \\
{\bar Q}_{nm} = 0 \quad {\rm for} \;\;\; n-m \geq 2 \;\; ,\;\; m-n \geq p_1
\phanta    \lab{param-2}
\er

Let us also introduce special notations for the first evolution parameters
$t_1 ,{\ti t}_1$ which in the sequel will be considered as space coordinates,
{\sl i.e.}, ${\ti t}_1 \equiv x$ and $t_1 \equiv y$. The lattice equations
of motion \rf{L-3} with $s=1$ imply that\foot{Arbitrary integration constants
are ignored without loss of generality.} :

(a) all matrix elements of ${\bar Q}$ can be expressed as functionals (w.r.t.
$x$) of $R_{n+1}, b_0 (n),\ldots ,\\
b_{p_1 -2}(n)$ at a {\em fixed} lattice site $n$;

(b) all matrix elements of $Q$ are explicitly expressed through
$R_{n+1}, b_0 (n),\ldots ,b_{p_1 -2}(n)$ via the formula \ct{enjoy} :
\be
Q_{(-)} = \({\bar Q}^{p_2 -1}\)_{(-)} + \( {1\over g}x\) \one   \lab{2-3-x}
\ee
where the last term is due to matching with the ``string'' equation
\rf{string-eq}.

There is a complete duality under interchanging
$\, p_1 \longleftrightarrow p_2\,$ of the orders of the matrix potentials
in \rf{2-1}, supplemented with interchanging
$x \equiv {\ti t}_1 \longleftrightarrow t_1 \equiv y \; ,\;
Q_{(-)} \longleftrightarrow {\bar Q}_{(+)}$ \ct{enjoy}. Namely, we have:

(a) all matrix elements of $Q$ can be expressed as functionals (w.r.t.
$y$) of $a_0 (n),\ldots ,a_{p_2 -1}(n)$ at a {\em fixed} lattice site
$n$;

(b) all matrix elements of ${\bar Q}$ are explicitly expressed through
$a_0 (n),\ldots ,a_{p_2 -1}(n)$ as:
\be
{\bar Q}_{(+)} = \( Q^{p_1 -1}\)_{(+)} + \( {1\over g}y\) \one  \quad ,\quad
R_{n+1} = Q^{p_1 -1}_{n+1,n} + {1\over g}n   \lab{2-3-y}
\ee
where the terms involving the two-matrix model coupling parameter $g$ come
{}from matching with the ``string'' equation \rf{string-eq}.

Using the parametrization \rf{param-1}--\rf{param-2}, the equations of the
auxiliary linear Lax problem \rf{L-1},\rf{L-2} acquire the form:
\br
\l \psi_n &=& \psi_{n+1} + a_0 (n) \psi_n + \sum_{k=1}^{p_2 -1}a_k
(n)\psi_{n-k}
\lab{7-1} \\
-{1\over g}\partder{}{\l} \psi_n &=& R_n \psi_{n-1} + b_0 (n) \psi_n +
\sum_{k=1}^{p_1 -1} \frac{b_k (n)}{R_{n+1} \ldots R_{n+k}} \psi_{n+k}
\lab{7-2} \\
\pa_x \psi_n &=& - R_n \psi_{n-1} \quad , \quad
\pa_y \psi_n = \psi_{n+1} + a_0 (n) \psi_n    \lab{7-3}
\er
Comparing eqs.\rf{7-1}--\rf{7-3} with \rf{spectr}, one identifies the
two-matrix model as a special constrained Toda lattice hierarchy.
Using \rf{7-3}, the implications of the lattice equations of motion:
\br
a_0 (n+k) \eq a_0 (n) + \pa_y \ln \( R_{n+1}\ldots R_{n+k-1}\) \; ,\;
a_0 (n-k) = a_0 (n) - \pa_y \ln \( R_{n}\ldots R_{n-k+1}\) \nonu\\
b_0 (n+k) \eq b_0 (n) + \pa_x \ln \( R_{n+1}\ldots R_{n+k-1}\) \; ,\;
b_0 (n-k) = b_0 (n) - \pa_x \ln \( R_{n}\ldots R_{n-k+1}\) \nonu
\er
and the string equation solutions \rf{2-3-x}--\rf{2-3-y}, one can rewrite
\rf{7-1}--\rf{7-3} and their compatibility conditions as continuum Lax
problem at a fixed lattice site $n$ :
\br
\l \psi_n = L(n)\psi_n \; ,\quad  - {1\over g}\partder{}{\l} \psi_n =
{\bar L}(n) \psi_n \; ,\quad \partder{}{t_r}\psi_n = \cL_r (n) \psi_n
\; ,\quad \partder{}{{\ti t}_s}\psi_n = - {\bar \cL}_s (n) \psi_n
\lab{L-cont} \\
\partder{}{t_r} L(n) = \Sbr{{\cL}_r (n)}{L(n)} \quad ,\quad
\partder{}{{\ti t}_s} L(n) = \Sbr{L(n)}{{\bar \cL}_s (n)}  \phanta
\lab{L-cont-1} \\
\partder{}{t_r} {\bar L}(n) = \Sbr{{\cL}_r (n)}{{\bar L}(n)} \quad ,\quad
\partder{}{{\ti t}_s} {\bar L}(n) = \Sbr{{\bar L}(n)}{{\bar \cL}_s (n)}
\phanta \lab{L-cont-2}\\
\Sbr{L(n)}{{\bar L}(n)} = {1\over g}\one \phantb   \lab{string-eq-cont}
\er
The explicit form of the continuum Lax operators depends on which equation
in \rf{7-3} we are using to express $\psi_{n \pm \ell}$ at neighboring sites
in terms of $\psi_n$. In the ``$x$-picture'' ({\sl i.e.}, using the first
eq.\rf{7-3}) we have:
\br
L(n) &\equiv& -D_x^{-1}\, R_{n+1} + a_0 (n) + \sum_{k=1}^{p_2 -1}
\frac{(-1)^k {\bar Q}^{p_2 -1}_{n,n-k}}{R_n\ldots R_{n-k+1}}
\Bigl( D_x - \pa_x \ln \( R_n \ldots R_{n-k+2}\)\Bigr)
\times \ldots \nonu\\
&\times& \Bigl( D_x - \pa_x \ln R_n \Bigr)\,D_x \lab{Ln-x} \\
{\bar L}(n) &\equiv& - D_x + b_0 (n) +
\sum_{k=1}^{p_1 -1} (-1)^k b_k (n) \Bigl( D_x + \pa_x \ln \( R_{n+1} \ldots
R_{n+k}\)\Bigr)^{-1} \times \ldots \nonu \\
&\times&  \Bigl( D_x + \pa_x\ln R_{n+1}\Bigr)^{-1}   \lab{Lnb-x} \\
{\bar \cL}_s (n) &\equiv& \sum_{k=1}^s \frac{(-1)^k {\bar Q}^{s}_{n,n-k}}
{R_n\ldots R_{n-k+1}} \Bigl( D_x - \pa_x \ln \( R_n \ldots R_{n-k+2}\)\Bigr)
\ldots D_x  \lab{Ls-x}  \qquad \\
\cL_r (n) &\equiv& Q^r_{nn} + \sum_{k=1}^r (-1)^k Q^r_{n,n+k}
R_{n+1}\ldots R_{n+k} \times \nonu \\
&\times &\Bigl( D_x + \pa_x \ln \( R_{n+1} \ldots
R_{n+k}\)\Bigr)^{-1} \ldots \Bigl( D_x + \pa_x\ln R_{n+1}\Bigr)^{-1}
\lab{Lr-x}
\er
where all coefficients are expressed in terms of $R_{n+1}, b_0 (n),\ldots ,
b_{p_1 -2}(n)$ at a fixed site $n$ through the lattice equations of motion
and \rf{2-3-x}. In the ``$y$-picture'' ({\sl i.e.}, using the second
eq.\rf{7-3}) we have:
\br
L(n) &\equiv& D_y + \sum_{k=1}^{p_2 -1} a_k (n) \Bigl( D_y - a_0 (n)
+ \pa_y \ln \( R_n \ldots R_{n-k+1}\)\Bigr)^{-1} \times \ldots \nonu \\
&\times&  \Bigl( D_y - a_0 (n) + \pa_y\ln R_n\Bigr)^{-1}   \lab{Ln-y} \\
{\bar L}(n) &\equiv& \( D_y - a_0 (n)\)^{-1} R_n + b_0 (n) \phantb \nonu \\
&+& \sum_{k=1}^{p_1 -1} Q^{p_1 -1}_{n,n+k}
\Bigl( D_y - a_0 (n) - \pa_y \ln \( R_{n+1} \ldots R_{n+k-1}\)\Bigr)
\ldots \Bigl( D_y - a_0 (n)\Bigr)  \lab{Lnb-y} \\
&= &\( D_y - a_0 (n)\)^{-1} R_n + \sum_{k=0}^{p_1 -3}
\cF^{p_1 -1}_k \( a_0 ,\ldots a_{p_1 -1}\) D_y^k + D_y^{p_1 -1}
\lab{Lnb-yy}\\
\!\!\!{\bar \cL}_s (n) \!\!\!&\equiv&\!\! \!
\sum_{k=1}^s {\bar Q}^s_{n,n-k} \Bigl( D_y - a_0 (n) +
\pa_y \ln \( R_n \ldots R_{n-k+1}\)\Bigr)^{-1}
\ldots \Bigl( D_y - a_0 (n) + \pa_y\ln R_n \Bigr)^{-1} \phantom{aaaaa}
\lab{Ls-y} \\
\cL_r (n) \!\!&\equiv&\!\!  Q^r_{nn} + \sum_{k=1}^r Q^r_{n,n+k}
\Bigl( D_y - a_0 (n) - \pa_y \ln \( R_{n+1} \ldots R_{n+k-1}\)\Bigr)
\ldots \Bigl( D_y - a_0 (n)\Bigr)    \lab{Lr-y}
\er
where all coefficients are expressed in terms of $a_0 (n),\ldots ,
a_{p_2 -1}(n)$ at a fixed site $n$ through the lattice equations of motion
and \rf{2-3-y}. In particular, they imply the form \rf{Lnb-yy} of the second
Lax operator ${\bar L}(n)$.
In \rf{Ln-x}--\rf{Lr-y} we have used notations $D_x,\, D_y$ for differential
operators to distinguish them from ordinary derivatives on functions.

Eqs.\rf{L-cont}--\rf{Lr-y} are the continuum analogs of the constrained Toda
lattice Lax equations \rf{L-1}--\rf{string-eq} without taking any continuum
(double-scaling) limit. Let us particularly stress that they explicitly
incorporate the whole information from the matrix-model string equation
\rf{string-eq} through \rf{2-3-x} and \rf{2-3-y} which were used in their
derivation. Their respective flows \rf{L-cont-1} and \rf{L-cont-2} are
compatible (commuting), so it is sufficient to consider only one of them.

In ref.\ct{enjoy} we used the ``$x$-picture'' Lax operators
\rf{Ln-x}--\rf{Lnb-x} for $p_2 =3$ and arbitrary $p_1$ to show that
the corresponding constrained Toda lattice hierarchy
\rf{L-1}--\rf{string-eq} is equivalent to the
$SL(3,1)$-KdV hierarchy defined by:
\be
L(n)= -D_x^{-1} R_{n+1} + 2b_1 (n) + b^2_0 (n) - \pa_x b_0 (n)
- 2b_0 (n) D_x + D_x^2  \lab{Lax-3}
\ee
The Lax operator \rf{Lax-3} can be cast into the form \rf{laxino} via a
simple gauge (similarity) transformation.
Now both $p_1$ and $p_2$ are arbitrary and let us assume for definiteness
$p_1 \leq p_2$.
Here, in order to conform with the formalism of the previous sections,
we shall use the ``$y$-picture'' Lax operators \rf{Ln-y} and \rf{Lnb-yy}.
Comparing $L(n)$ \rf{Ln-y} with \rf{3-3} we see that the former
is a {\em constrained}~ $2(p_2 -1)$-boson KP Lax operator with coefficients:
\be
A^{(p_2 -1)}_{p_2 -k}= a_k (n) \quad ,\quad
B^{(p_2 -1)}_{p_2 -k}= a_0 (n) - \pa_y \ln \Bigl( R_n \bigl(\{ a\}\bigr)
\ldots R_{n-k+1}\bigl(\{ a\}\bigr)\Bigr)  \lab{ABab}
\ee
which depend on
$p_2$ fields only: $\{ a\} \equiv \( a_0 (n),\ldots ,a_{p_2 -1}(n) \)$.
Thus, there is only one independent current, {\sl e.g.},
$B^{(p_2 -1)}_{p_2 -1}$.

On the other hand, the second Lax operator ${\bar L}(n)$
\rf{Lnb-y}--\rf{Lnb-yy} has precisely the form of a $SL(p_1 ,1)$-KdV Lax
operator, cf. \rf{1ff} for $p_1 = M+1$. Moreover, when $p_1 \leq p_2$ its
coefficients depend only
on $a_0 (n),\ldots ,a_{p_1 -1}(n)$ which easily follows from \rf{2-3-y}
and the explicit parametrization \rf{param-1}--\rf{param-2}. Thus, the
constrained generalized Toda lattice hierarchy \rf{L-1}--\rf{string-eq},
describing the two-matrix string model \rf{2-1}, is equivalent to the
$SL(p_1 ,1)$-KdV hierarchy \rf{L-cont-2} with finite number of flows.

\sect{Conclusions}
We conclude with a list of remarks.
\begin{itemize}
\item
We have described here the process of reduction of the multi-boson
KP hierarchies analyzed in two special settings, or gauges, of the associated
Toda matrix spectral problem.
Our basic gauge was the Drinfeld-Sokolov gauge naturally associated with the
Toda lattice hierarchy.
Another gauge, which proved to be useful in our discussion, especially
when dealing with the Poisson bracket structure, was the diagonal gauge
related to the Volterra lattice.
Modes associated with the Volterra lattice abelianized and simplified the
analysis of the second Hamiltonian structure.
These two gauges led to two different Lax formulations having different
transformation properties under the residual gauge transformations.
\item
We have obtained in this paper a coherent approach to describe
generalized KP-KdV ({\sl i.e.}, $SL(M+1,M-k)$-KdV)
hierarchies and their Poisson bracket structures.
Variables (free currents) abelianizing the second bracket structure were
instrumental tools in this analysis. They naturally lead to the appearance
of the graded $SL(M+1,M-k)$ Kac-Moody algebras.
This raises the question about the origin of the graded algebra in this
setting. One possible explanation could be given in terms of the underlying
lattice structure when one recalls that the transition from Toda to Volterra
lattice involves separation in even and odd sites.
This could be an intuitive way of seeing how the grading could have been
introduced into the formalism.
\item
The free fields (currents) abelianizing the second Hamiltonian structure
of the KP-KdV hierarchies bring about another noticeable
result. Namely, they yield explicit free-field representations of the
nonlinear $W(M,M-k)$ algebras, isomorphic to the $SL(M+1,M-k)$-KdV
Poisson bracket algebras, which generalize Zamolodchikov's nonlinear
$W_{M+1}$ algebra.
\item
Lattice translations in the underlying Toda and Toda-like hierarchies
naturally give rise
to similarity transformations of the corresponding KP-KdV Lax operators
which, first, preserve the Hamiltonians and the Poisson structures, and
second, systematically generate the pertinent auto-B\"{a}cklund
transformations for the generalized $SL(M+1,M-k)$-KdV hierarchies.
\item
The physical relevance of the structures, defined by the reduction process
described in this paper, is now strongly enhanced by the natural appearance
of the generalized $SL(M+1,1)$-KdV hierarchy within the context of the
two-matrix string model.
More precisely, ref.\ct{enjoy} identified the two-matrix model in the
simplest nontrivial case with a coupled system of $2+1$-dimensional KP
and modified KP ($(m)KP_{2+1}$) integrable equations subject to a specific
``symmetry'' constraint.
This constraint together with the Miura-Konopelchenko map \ct{Konop} for
$(m)KP_{2+1}$ are the images in the continuum of the matrix-model
string equation \rf{string-eq}.
In particular, the two-matrix model susceptibility is a solution to the above
string-constrained $KP_{2+1}$ equation.
The string-constrained $(m)KP_{2+1}$ system was shown to be equivalent to the
$1+1$-dimensional generalized $SL(3,1)$-KdV hierarchy \rf{Lax-3},\rf{laxino}.
Previously, the generalized KP-KdV models were obtained in \ct{BX9311a}
{}from two-matrix models by imposing {\em ad hoc} additional Dirac
constraints on the multi-boson KP hierarchy.
\end{itemize}

\lskip
{\bf Acknowledgements.} A.H.Z. thanks UIC for hospitality during his stay in
Chicago and FAPESP for financial support. H.A. acknowledges IFT-UNESP
for hospitality during completion of this paper and FAPESP for financial
support. The authors are grateful to J.F. Gomes for discussions.

\end{document}